\documentclass[10pt]{article}

\usepackage{rotating}  
\usepackage{theorem}   
\usepackage{bm}  
\usepackage{amsmath,amsfonts,amssymb}
\usepackage{booktabs}  
\usepackage{pdfsync}   
\usepackage{adjustbox}
\usepackage{amsmath,amssymb,amsfonts}
\usepackage{algorithmic}
\usepackage[]{algorithm2e}
\usepackage{placeins}
\usepackage{afterpage}
\usepackage{multicol}
\usepackage{comment}
\usepackage{floatrow}
\usepackage{lipsum}
\usepackage[utf8]{inputenc}
\usepackage[english]{babel}
\usepackage{csquotes}
\usepackage{rotating}
\usepackage{graphicx}
\usepackage{url}
\usepackage{amsmath}
\usepackage{mathtools}
\usepackage{algorithm2e}

\usepackage{subcaption}
\usepackage{rotating}  
\usepackage{theorem}   
\usepackage{bm}  
\usepackage{amsmath,amsfonts,amssymb,mathabx}
\usepackage{booktabs}  
\usepackage{pdfsync}   
\usepackage{pdfpages}
\usepackage{float}
\usepackage{threeparttable}
\usepackage{makecell}
\usepackage{adjustbox}
\usepackage{listings}

\usepackage[explicit]{titlesec}
\usepackage{hyphenat}
\usepackage{ragged2e}
\RaggedRight

\usepackage{algorithmic}
\usepackage{multicol}

\usepackage{rotating}  
\usepackage{theorem}   
\usepackage{bm}  
\usepackage{amsmath,amsfonts,amssymb}
\usepackage{booktabs}  
\usepackage{pdfsync}   
\usepackage{graphicx}
\usepackage{latexsym}
\usepackage{amsmath}
\usepackage{amssymb}
\usepackage{fancyvrb}
\usepackage{colortbl}
\usepackage{enumerate}
\usepackage{pifont}
\usepackage{stmaryrd}
\usepackage{textcomp}
\usepackage{fncylab}
\usepackage{multirow}
\usepackage{paralist}
\usepackage{wrapfig}
\usepackage{longtable}
\usepackage{lscape}
\usepackage{hyperref}
\usepackage{lipsum}

\usepackage{garamondx}
\usepackage[T1]{fontenc}
\usepackage{amsmath}
\usepackage{graphicx}
\usepackage{xcolor}

\usepackage{geometry}
\usepackage{fancyhdr}
\usepackage[labelfont={footnotesize,bf} , textfont=footnotesize]{caption}
\makeatletter
\renewcommand\tagform@[1]{\maketag@@@ {\ignorespaces {\footnotesize{\textbf{Equation}}} #1.\unskip \@@italiccorr }}
\makeatother
\titleformat{\section}
  {\normalfont}{\thesection}{1em}{\MakeUppercase{\textbf{#1}}}
\titlespacing\section{0pt}{0pt}{-10pt}
\titleformat{\subsection}
  {\normalfont}{\thesubsection}{1em}{\textit{#1}}
\titlespacing\subsection{0pt}{0pt}{-8pt}

\makeatletter
\newcommand\sixteen{\@setfontsize\sixteen{17pt}{6}}
\renewcommand{\maketitle}{\bgroup\setlength{\parindent}{0pt}
\begin{flushleft}
\sixteen\bfseries \@title
\medskip
\end{flushleft}
\textit{\@author}
\egroup}
\makeatother

\usepackage[sort&compress]{natbib}
\makeatletter
\renewcommand\@biblabel[1]{\textbf{#1.}\hfill}
\makeatother

\bibpunct{}{}{,~}{s}{,}{,}
\colorlet{punct}{red!60!black}
\definecolor{background}{HTML}{EEEEEE}
\definecolor{delim}{RGB}{20,105,176}
\colorlet{numb}{magenta!60!black}

\lstdefinelanguage{json}{
    basicstyle=\normalfont\ttfamily,
    numbers=left,
    numberstyle=\scriptsize,
    stepnumber=1,
    numbersep=8pt,
    showstringspaces=false,
    breaklines=true,
    frame=lines,
    backgroundcolor=\color{background},
    literate=
     *{0}{{{\color{numb}0}}}{1}
      {1}{{{\color{numb}1}}}{1}
      {2}{{{\color{numb}2}}}{1}
      {3}{{{\color{numb}3}}}{1}
      {4}{{{\color{numb}4}}}{1}
      {5}{{{\color{numb}5}}}{1}
      {6}{{{\color{numb}6}}}{1}
      {7}{{{\color{numb}7}}}{1}
      {8}{{{\color{numb}8}}}{1}
      {9}{{{\color{numb}9}}}{1}
      {:}{{{\color{punct}{:}}}}{1}
      {,}{{{\color{punct}{,}}}}{1}
      {\{}{{{\color{delim}{\{}}}}{1}
      {\}}{{{\color{delim}{\}}}}}{1}
      {[}{{{\color{delim}{[}}}}{1}
      {]}{{{\color{delim}{]}}}}{1},
}

\title{The Impact of Generative AI on Student Churn and the Future of Formal Education}

\author{
Stephen Elbourn*$^{a}$ \\ \medskip
$^{a}$Macquarie University, Sydney, Australia \\  \medskip
stephen.elbourn@students.mq.edu.au
}

\begin{document}

\vspace*{.01 in}
\maketitle
\vspace{.12 in}

\section*{abstract}

In the contemporary educational landscape, the advent of Generative Artificial Intelligence (AI) presents unprecedented opportunities for personalised learning, fundamentally challenging the traditional paradigms of education. This research explores the emerging trend where high school students, empowered by tailored educational experiences provided by Generative AI, opt to forgo traditional university degrees to pursue entrepreneurial ventures at a younger age. To understand and predict the future of education in the age of Generative AI, we employ a comprehensive methodology to analyse social media data.
Our approach includes sentiment analysis to gauge public opinion, topic modelling to identify key themes and emerging trends, and user demographic analysis to understand the engagement of different age groups and regions. We also perform influencer analysis to identify key figures shaping the discourse and engagement metrics to measure the level of interest and interaction with AI-related educational content. Content analysis helps us to determine the types of content being shared and the prevalent narratives, while hashtag analysis reveals the connectivity of discussions. The temporal analysis tracks changes over time and identifies event-based spikes in discussions.
The insights derived from this analysis include the acceptance and adoption of Generative AI in education, its impact on traditional education models, the influence on students' entrepreneurial ambitions, and the educational outcomes associated with AI-driven personalised learning. Additionally, we explore public sentiment towards policies and regulations and use predictive modelling to forecast future trends. This comprehensive social media analysis provides a nuanced understanding of the evolving educational landscape, offering valuable perspectives on the role of Generative AI in shaping the future of education.

\section*{keywords}
Generative Artificial Intelligence (AI), Personalised Learning, Education Transformation, Entrepreneurial Ambitions, Social Media Analysis.

\vspace{.12 in}


\section{introduction}

In this Section, we begin with an overview of the work and explain the problem statement: understanding and predicting the impact of Generative Artificial Intelligence (AI) on the future of education. Next, we provide our contributions and explore the emerging trend where high school students, empowered by tailored educational experiences provided by Generative AI, opt to forgo traditional university degrees to pursue entrepreneurial ventures at a younger age. Finally, we outline the organisation of this paper.

\subsection{Overview and Problem Statement}

In recent years, the field of education has witnessed significant transformation driven by technological advancements~\cite{farhood2024evaluating}. Among these advancements, Generative Artificial Intelligence (AI) has emerged as a powerful tool capable of reshaping traditional educational paradigms. Generative AI, which refers to AI systems that can create new content and generate personalised experiences, has found diverse applications across various domains, including education~\cite{beheshti2023empowering}. These AI systems can produce tailored learning materials, provide personalised feedback, and adapt to individual learning styles, thus offering unprecedented opportunity for personalised learning~\cite{beheshti2020towards}.

Traditional education systems, characterised by standardised curricula and uniform teaching methods, have often struggled to meet students' diverse needs. In contrast, Generative AI enables the creation of customised educational experiences that cater to individual student's unique learning preferences and paces~\cite{brender2024s}. This personalised approach can potentially enhance student engagement, improve learning outcomes, and foster a more inclusive learning environment.

One of the most profound impacts of Generative AI in education is its influence on students' career choices and aspirations~\cite{abc2024chatgpt}. With access to highly personalised learning experiences, many high school students are increasingly opting to bypass traditional university education in favour of entrepreneurial ventures. This trend is driven by the belief that personalised learning equips them with the skills and knowledge necessary to succeed in a rapidly changing job market. As a result, there is a growing movement of young entrepreneurs leveraging their AI-enhanced education to innovate and create new business opportunities at an earlier age~\cite{australianAIframework2024}.

Social media platforms serve as a rich data source for understanding these emerging trends~\cite{beheshti2022social}. They provide a real-time and unfiltered view of public opinions, experiences, and discussions related to education and technology. By analysing social media data, researchers can gain valuable insights into how Generative AI is perceived and utilised in an educational context. This analysis can reveal patterns, sentiments, and trends critical for predicting the future of education in the age of Generative AI.

Despite the potential benefits, integrating Generative AI in education raises many challenges and concerns. Issues relating to data privacy, algorithmic bias, and the digital divide must be addressed to ensure equitable access to AI-driven educational resources. Furthermore, the shift towards personalised learning and entrepreneurial pursuits requires reevaluating traditional educational metrics and success indicators.

\textbf{Research Problem.}
In this paper, we focus on the problem of understanding and predicting the impact of Generative Artificial Intelligence (AI) on the future of education, particularly focusing on the emerging trend where high school students, empowered by AI-driven personalised learning experiences, are increasingly opting to forgo traditional university degrees in favour of entrepreneurial ventures at a younger age. This research explores how Generative AI is reshaping educational paradigms, influencing career choices, and altering the landscape of higher education and entrepreneurship. Through comprehensive social media analysis, the study aims to uncover public sentiments, trends, and behaviours related to AI-driven education and its implications for the future.

This research aims to explore these dynamics by conducting a comprehensive analysis of social media discussions on Generative AI and education. By leveraging advanced analytical techniques, we seek to understand the current landscape, predict future trends, and provide actionable insights for educators, policymakers, and stakeholders. This background sets the stage for a detailed examination of how Generative AI is transforming education and what this means for the future of learning and career development.

\subsection{Contributions}

This paper contributes to the field of educational technology and the study of Generative Artificial Intelligence (AI) in education. In the following we present the main contributions of this paper.

\subsubsection{Contextualisation of Social Media Data using ProcessGPT}

This paper introduces an innovative method for automatically contextualising social media data by leveraging ProcessGPT~\cite{beheshti2023processgpt}. ProcessGPT utilises advanced natural language processing techniques to interpret and organise social media content, enhancing the understanding of its context and relevance to educational settings. This approach not only improves the accuracy and relevance of data analysis but also provides a framework for integrating diverse data sources, thereby enriching the educational research landscape with more comprehensive and context-aware insights.

\subsubsection{Learning Distributed Representations and Deep Embedded Clustering}

Building on the contextualised social data, this paper implements a novel approach to learn distributed representations and perform deep embedded clustering of texts. This technique employs state-of-the-art machine learning algorithms to transform the contextualised data into meaningful representations, which are then used for clustering. By uncovering underlying patterns and insights from the processed data, this approach facilitates a deeper understanding of the impact of Generative AI on the future of formal education. The clustering results enable educators and researchers to identify trends, categorise information effectively, and develop targeted interventions to enhance educational outcomes. This contribution underscores the potential of Generative AI to revolutionise educational technology by providing data-driven solutions that are both adaptive and insightful.



\subsection{Summary and Outline}


In this Section, we discussed a broad picture of the problem and covered the motivation, problem statement, and contributions of this paper.
The remainder of this paper is structured in the following manner:

\begin{itemize}

\item
Section 2 will present the background and analyse the related work
in the education domain, Social data analytics, and Generative AI. We will review the current state-of-the-art approaches, including Generative AI in Education, Impacts of AI on Educational Outcomes, Shift in Educational and Career Choices, Social Media Analysis in Educational Research, Ethical and Societal Implications of AI in Education, Policy and Regulatory Perspectives, Future Trends and Predictions. We will provide a summary to compare the methods and techniques reviewed in the four areas.

\item
Section 3 will delve into our comprehensive methodology, starting with the Social Media Analysis Framework, which encompasses sentiment analysis, topic modelling, user demographic analysis, influencer analysis, engagement metrics, content analysis, hashtag analysis, and temporal analysis. This framework is designed to systematically extract and interpret data from social media platforms, providing a robust foundation for understanding public discourse and sentiment regarding Generative AI in education. We will then detail our approach to identifying emerging trends, focusing on the dynamic shifts in educational practices and in career aspirations influenced by AI-driven personalised learning. 

\item
In Section~4, we will present the experiment, along with its results and evaluation. We will begin with a motivating scenario to help clarify our approach, followed by a detailed explanation of the dataset and experimental setup. Finally, we will conclude this section with an in-depth discussion of the evaluation results.

\item
Finally, in Section~5, we will conclude the study by summarizing the proposed method and discussing potential future directions that build upon our findings.
\end{itemize}


\section{Background and State-of-the-Art}

In this section, we present a comprehensive review of the related work in the domains of education, social data analytics, and Generative AI. This section aims to provide a contextual foundation for our research on understanding and predicting the impact of Generative AI on the future of education. We review current state-of-the-art approaches and research, focusing on the following areas: data curation in social media analytics, generative AI in education, impacts of AI on educational outcomes, shift in educational and career choices, ethical and societal implications of AI in education, and future trends and predictions. We conclude with a summary comparing the methods and techniques reviewed.

\subsection{Data Curation in Social Media Analytics}

Data curation is a critical step in transforming raw social data into meaningful and actionable insights. It involves processes such as identifying relevant data sources, extracting and cleaning data, enriching and linking data to domain knowledge, and managing the curated data over time~\cite{beheshti2019datasynapse}. Social data generated by platforms like Twitter (recently rebranded as X)\footnote{https://x.com/}, Facebook\footnote{https://www.facebook.com/}, and LinkedIn\footnote{https://www.linkedin.com/} presents unique challenges due to its unstructured nature, high volume, and need for real-time processing. Traditional data processing methods are often insufficient for handling this complexity and scale. Feature extraction~\cite{li2021comprehensive}, data enrichment, and contextualisation are key challenges in social data curation, requiring advanced methods to make raw data interpretable and useful~\cite{aldoseri2023re}.

Several related works in data curation contribute to this field, including using Knowledge Graphs like Wikidata\footnote{https://www.wikidata.org/} and DBpedia\footnote{https://www.dbpedia.org/} for enriching and annotating raw data, and data integration techniques in ETL~\cite{vassiliadis2009taxonomy} systems focusing on schema integration and entity deduplication. Crowdsourcing techniques have also been employed for keyword extraction and named entity recognition in social media content~\cite{beheshti2018crowdcorrect}. Sentiment analysis and classification algorithms are widely used to analyse sentiments and classify social media posts~\cite{mouthami2013sentiment}.

DataSynapse~\cite{beheshti2019datasynapse} is a notable framework addressing these challenges by providing scalable solutions for transforming raw social data into contextualised and curated data. It introduces the concepts of Semantic-Item and Featurised-Item to enable customisable feature-based extraction and enrichment from diverse data sources~\cite{khadivizand2020towards}. Additionally, it supports linking extracted data to domain knowledge through Contextualised-Item, promoting data transformation into actionable knowledge. DataSynapse also implements scalable algorithms for this transformation and introduces the notion of a Knowledge Lake, a centralised repository containing both raw and contextualised data for big data analytics~\cite{rajabi2016interlinking}.

DataSynapse provides an extensible and scalable microservice-based architecture supporting various data curation tasks, such as extracting, linking, merging, and summarising data. It has been evaluated in practical scenarios, demonstrating significant improvements in the quality of extracted knowledge compared to traditional curation pipelines. For instance, it has been used to analyse tweets related to government budgets and identify social issues such as health and public safety. By linking extracted entities from tweets to domain knowledge bases like the Australian Government's Budget Knowledge Base, DataSynapse enables more precise and contextually relevant insights. Future directions in data curation research aim to enhance the precision and recall of curation techniques, improve scalability, and develop new methods for intelligent narrative discovery and summarisation. DataSynapse's approach and methodologies pave the way for more effective social media analytics, providing a robust foundation for ongoing and future research in data curation~\cite{beheshti2022social}.

\subsection{Generative AI in Education}

Artificial Intelligence (AI) refers to the ability of machines to perform tasks that typically require human intelligence, such as learning, reasoning, problem-solving, and adapting to new situations~\cite{nilsson2009quest}. AI systems can be broadly classified into three primary types based on the components of human intelligence they emulate~\cite{beheshti2023empowering}: Analytical AI, Cognitive AI, and Generative AI. Each type plays a distinct role in acquiring and applying knowledge and skills, influenced by knowledge, experience, and creativity~\cite{shabani2023rule}.

Analytical AI focuses on understanding raw data and transforming it into contextualised data and knowledge~\cite{beheshti2023empowering,shang2019data}. It employs data mining, statistical analysis, machine learning, and deep learning techniques to extract meaningful insights from large datasets. These systems are commonly used in tasks like pattern recognition, anomaly detection, and predictive analytics. Analytical AI is foundational in fields where data-driven decision-making is crucial, providing the backbone for understanding and interpreting vast amounts of data.

Cognitive AI aims to mimic human cognitive processes, such as perception, memory, and reasoning, to facilitate decision-making~\cite{beheshti2023empowering,zhao2022emotion}. It focuses on utilising knowledge and experience to annotate and enrich data, making it valuable for healthcare, finance, and education applications. Cognitive AI systems assist knowledge workers by providing insights and recommendations based on past experiences and learned knowledge, thereby enhancing decision-making processes.

Generative AI, on the other hand, delves into the neural mechanisms involved in creative thinking and problem-solving~\cite{beheshti2023empowering,brynjolfsson2023generative}. It focuses on generating new content, such as text, images, and processes, by leveraging techniques like deep learning, reinforcement learning, and evolutionary algorithms. Generative AI has found applications in various fields, including art, music, design, and business process management~\cite{epstein2023art}. It not only automates content creation but also improves processes by generating new models and instances based on learned patterns.

Integrating Generative AI with Knowledge Base 4.0 (KB~4.0~\cite{beheshti2022knowledge}) represents a significant advancement in AI research~\cite{beheshti2023empowering}. KB 4.0 is a robust data repository that links knowledge and experience, empowering Generative AI to not only generate new content but also create new processes when needed. The architecture of KB 4.0 includes components such as Data Lake Services, Knowledge Lake Services, Cognitive AI Services, and Generative AI Services~\cite{beheshti2022knowledge,shabani2024comprehensive,wang2022performance}. These components work together to construct and maintain a comprehensive Knowledge Graph~\cite{beheshti2016scalable}, enabling the continuous enrichment and contextualisation of data.

The potential of Generative AI extends beyond simple content generation to revolutionising business processes~\cite{beheshti2023processgpt,jafari2024empowering,jafari2022automatic}. For example, in the education domain, Generative AI can help new teachers identify creativity patterns and develop lesson plans that promote creative thinking among students~\cite{su2023unlocking}. Similarly, in the healthcare sector, Generative AI can assist novice practitioners in making informed diagnoses by analysing patient data and generating possible treatment plans~\cite{shokrollahi2023comprehensive}. Integrating KB 4.0 with Generative AI ensures that these systems have access to a rich source of contextualised knowledge, allowing for continuous improvement and innovation.

Previous research has focused on developing AI models that generate educational content, such as problem sets, quizzes, and interactive simulations~\cite{wang2023learning}. For example, adaptive learning platforms like Carnegie Learning~\cite{carnegielearning2024} and Knewton~\cite{wiley2024} utilise AI to personalise educational content based on student performance and learning preferences. These systems employ natural language processing, machine learning, and reinforcement learning to generate content that adaptively meets individual student needs.

\subsection{Impacts of AI on Educational Outcomes}

The integration of AI in education has the potential to significantly impact educational outcomes. Research has shown that AI-enabled personalised learning can improve academic performance, retention rates, and motivation among students~\cite{zhai2021review}. For instance, studies by Kulik and Fletcher (2016)~\cite{kulik2016effectiveness} demonstrated that adaptive learning technologies could improve student achievement by tailoring instruction to individual learning needs.
Moreover, AI can provide real-time feedback and assessment, allowing educators to identify and address learning gaps promptly. The use of AI in formative assessment~\cite{bennett2011formative}, Hattie and Timperley (2007)~\cite{hattie2007power}, has been shown to enhance student learning by providing targeted feedback that guides improvement. Additionally, AI-driven analytics can help educators make data-informed decisions to optimise instructional strategies and resource allocation.

Beyond these direct educational benefits, AI's impact on educational outcomes can be observed in several other key areas. Firstly, AI technologies enable the development of intelligent tutoring systems (ITS)~\cite{nwana1990intelligent} that offer one-on-one instruction tailored to the individual student’s pace and learning style. Research by VanLehn (2011)~\cite{vanlehn2011relative} has shown that intelligent tutoring systems can be as effective as human tutors, significantly improving student learning in subjects like mathematics and science.

AI also supports the creation of adaptive learning~\cite{kerr2016adaptive} environments that adjust the difficulty and nature of tasks based on the student's performance. These environments can track student progress over time and provide customised resources to address specific weaknesses, thereby fostering a more personalised and effective learning experience. For example, the Knewton~\cite{wiley2024} adaptive learning platform uses algorithms to create personalised learning paths, enhancing both engagement and achievement among students.

In addition to supporting individual learning, AI can facilitate collaborative learning~\cite{smith1992collaborative} by forming intelligent groups of students based on their strengths and weaknesses. AI algorithms can analyse students' interaction patterns and suggest optimal groupings for collaborative projects, ensuring that each group is balanced and that members can learn from each other. This approach not only enhances individual learning outcomes but also develops students' teamwork and communication skills.

AI's ability to process and analyse large datasets also allows for a deeper understanding of educational trends and student behaviours. By analysing data from various sources, including learning management systems, online course platforms, and social media, AI can identify patterns and trends that might be missed by human analysts. For instance, AI can detect early signs of student disengagement or predict which students are at risk of dropping out, enabling timely interventions to improve retention rates~\cite{supangat2021churn}.

Furthermore, AI-driven educational tools can enhance accessibility for students with disabilities~\cite{eziamaka2024ai}. Tools such as speech recognition software, text-to-speech converters, and AI-powered translators can help students with visual, auditory, or language impairments access educational content more easily. These tools ensure that all students can benefit from personalised learning experiences, thereby promoting inclusivity in education.

The potential of AI to transform educational outcomes is also evident in higher education and professional training contexts. AI can support lifelong learning by providing personalised learning pathways that adapt to the evolving needs of learners throughout their careers~\cite{reddy2015personalized}. For example, AI-driven platforms like Coursera\footnote{https://www.coursera.org/} and edX\footnote{https://www.edx.org/} use machine learning algorithms to recommend courses and learning materials based on users' past activities and preferences, helping professionals stay updated with the latest industry trends and skills.

In research, AI can facilitate the discovery of new knowledge and the advancement of educational practices. By analysing vast amounts of research data, AI can identify emerging trends, gaps in the literature, and potential areas for further investigation~\cite{lee2024aid}. AI-driven literature review tools, such as Iris.ai\footnote{https://iris.ai/}, can assist researchers in navigating the ever-growing body of academic literature, making it easier to find relevant studies and synpapere findings.

Despite the numerous benefits, integrating AI into education also presents challenges and concerns~\cite{hwang2020vision}. Issues related to data privacy, algorithmic bias, and the digital divide must be addressed to ensure that the benefits of AI are equitably distributed~\cite{schiliro2022deepcog}. Ensuring that AI systems are transparent and that their decision-making processes can be understood by educators and students is crucial for building trust and acceptance~\cite{schmidt2020transparency}.

\subsection{Shift in Educational and Career Choices}

Generative AI's influence extends beyond academic performance to shaping students' career choices and aspirations. With access to personalised learning experiences, students are increasingly considering alternative educational pathways, such as entrepreneurship. This trend is particularly evident among high school students who, empowered by AI-driven learning, are opting to forgo traditional university education in favour of starting their ventures.

Research by Bakhshi et al. (2017)~\cite{bakhshi2017future} suggests that the skills acquired through personalised learning, such as critical thinking, problem-solving, and adaptability, are highly valued in the entrepreneurial ecosystem. Consequently, students perceive AI-enabled education as a means to gain relevant skills for the rapidly evolving job market. This shift poses challenges for traditional educational institutions, necessitating a reevaluation of curricula and success metrics to align with students' changing aspirations.

The impact of Generative AI on educational and career choices is multifaceted~\cite{holmes2023guidance}. AI-facilitated personalised learning experiences provide students with tailored educational content that matches their individual learning styles and paces. This personalisation fosters a sense of autonomy and motivation, encouraging students to explore and develop their unique talents and interests. As a result, students become more confident in their abilities and more willing to take risks, such as pursuing entrepreneurial ventures.

Generative AI tools enable students to gain hands-on experience in various fields by simulating real-world scenarios and providing interactive learning environments~\cite{salinas2024using}. For instance, AI-driven platforms can simulate business environments, allowing students to practice decision-making, financial management, and strategic planning. These simulations help students develop practical skills that are directly applicable to entrepreneurial activities.

Integrating AI in education also facilitates lifelong learning and continuous skill development. With the rapid pace of technological advancements, the job market constantly evolves, requiring individuals to update their skills regularly~\cite{hutson2023rethinking}. AI-powered educational platforms can provide ongoing learning opportunities, enabling individuals to acquire new skills and knowledge throughout their careers. This continuous learning model aligns with the needs of the modern workforce, where adaptability and lifelong learning are essential for success.

Furthermore, the rise of AI-enhanced education contributes to a shift in the perception of traditional educational pathways~\cite{wang2023preparing}. The conventional model of completing a university degree before entering the job market is being challenged by the availability of alternative educational resources. Online courses, boot camps, and AI-driven learning platforms offer flexible and affordable options for acquiring valuable skills. These alternatives particularly appeal to students who prefer a more practical, hands-on approach to learning.

The trend towards entrepreneurship among AI-empowered students is also influenced by the accessibility of digital tools and platforms that facilitate business creation and management~\cite{adithya2024entrepreneurship}. AI technologies can assist in various aspects of entrepreneurship, such as market research, product development, and marketing strategies. For example, AI algorithms can analyse market trends and consumer preferences, giving entrepreneurs insights that inform their business decisions. Additionally, AI-driven marketing tools can automate customer engagement and personalise marketing campaigns, increasing the efficiency and effectiveness of business operations~\cite{sharma2023ai}.

While the shift towards AI-driven personalised learning and entrepreneurship presents numerous opportunities, it also raises several challenges for educational institutions~\cite{jian2023personalized}. Traditional universities must adapt to the changing landscape by incorporating AI technologies into their curricula and offering flexible learning options. This may involve developing new courses focused on AI and entrepreneurship, as well as integrating AI tools into existing programs to enhance personalised learning and skill development.

Educational institutions must also consider AI's implications for educational equity and access. Ensuring that all students have access to AI-driven learning resources is crucial for preventing disparities in educational opportunities. Institutions should invest in the necessary infrastructure and support students facing barriers to accessing AI technologies.

\subsection{Ethical and Societal Implications of AI in Education}

Integrating Generative AI in education raises several ethical and societal concerns~\cite{alasadi2023generative}. Issues related to data privacy, algorithmic bias, and the digital divide must be addressed to ensure equitable access to AI-driven educational resources. There is a need for transparency and accountability in AI systems to prevent discriminatory practices and ensure fairness.

Moreover, the shift towards personalised learning and entrepreneurial pursuits requires reevaluating traditional educational metrics and success indicators. Ethical considerations must guide the development and implementation of AI technologies to avoid exacerbating existing inequalities and to promote inclusive and accessible education for all students~\cite{singh2024ethical}.

Policymakers and educational institutions play a crucial role in addressing the challenges of integrating Generative AI in education. Policy recommendations focus on ensuring data privacy, mitigating algorithmic bias, and providing equitable access to AI-driven educational resources~\cite{shahriar2023survey}. For example, the European Commission's guidelines on AI ethics stress the importance of human-centric AI, transparency, and accountability in educational applications~\cite{smuha2019eu}.

Educational institutions must also adapt to the changing landscape by revising curricula, investing in teacher training, and fostering collaborations with AI developers. Policies should encourage the responsible use of AI, promote digital literacy, and support research on the long-term impacts of AI in education~\cite{koravuna2020educational}.

\subsection{Future Trends and Predictions}

The future of education in the age of Generative AI holds promising possibilities. Emerging trends include the rise of AI-driven tutoring systems, virtual and augmented reality in education, and the use of AI for lifelong learning and skill development~\cite{mello2023education}. Predictive analytics and learning analytics are expected to play a significant role in personalised learning, helping educators anticipate student needs and tailor instruction accordingly~\cite{farhood2024evaluating}.

One significant trend is the development of AI-driven tutoring systems~\cite{baumgart2022knowledge}. These systems leverage Generative AI to provide personalised tutoring experiences, adapting content and feedback to each student's unique learning style and pace. Such systems can offer one-on-one tutoring at scale, making high-quality education accessible to a broader audience. AI tutors can assist with homework, explain complex concepts, and even prepare students for exams through tailored practice sessions.

Virtual and augmented reality (VR/AR) are also gaining traction as transformative educational tools~\cite{al2023analyzing}. These technologies create immersive learning environments that enhance engagement and retention. For instance, VR can simulate historical events, scientific phenomena, or complex machinery, allowing students to explore and interact with content in a way that traditional methods cannot match. AR can overlay digital information onto the physical world, enriching real-world learning experiences and providing immediate, context-aware assistance.

Lifelong learning and continuous skill development are becoming increasingly important in a rapidly changing job market~\cite{sobhanmanesh2023cognitive}. AI can facilitate this by providing personalised learning paths that adapt to an individual's career progression and skill requirements. Generative AI can create custom courses, recommend relevant learning resources, and even generate new educational content on demand~\cite{pesovski2024generative}. This ensures that learners can continuously update their skills and knowledge to remain competitive.

Predictive analytics and learning analytics are crucial in advancing personalised learning~\cite{khor2023systematic}. AI systems can identify patterns and predict future learning needs by analysing data on student performance, engagement, and learning behaviours. Educators can use these insights to tailor their instruction, provide timely interventions, and optimise curriculum design. This data-driven approach enables a more responsive and adaptive educational environment that caters to each student's needs.

Moreover, research by Luckin et al. (2016)~\cite{luckin2016intelligence} highlights the potential of AI to transform educational practices, making learning more adaptive, engaging, and inclusive. AI technologies can help create learning experiences that are not only effective but also enjoyable and motivating for students. Adaptive learning platforms can adjust the difficulty of tasks in real time, ensuring that students remain challenged but not overwhelmed.

However, realising this potential requires addressing several ethical, societal, and policy challenges. Data privacy is a significant concern, as AI systems often rely on extensive data collection to function effectively~\cite{jain2016big}. Ensuring that student data is protected and used responsibly is paramount. Algorithmic bias is another issue, where AI systems may inadvertently perpetuate existing inequalities or develop biased learning recommendations~\cite{baker2022algorithmic}. Developing fair and transparent AI systems is crucial to mitigate these risks.

The digital divide is a persistent challenge, with unequal access to technology and internet connectivity creating disparities in educational opportunities~\cite{cullen2001addressing}. Policymakers and educational institutions must work together to ensure that AI-driven educational tools are accessible to all students, regardless of their socioeconomic background. This may involve investing in infrastructure, providing affordable devices, and offering digital literacy training.

Finally, integrating Generative AI into education requires reevaluating traditional educational metrics and success indicators. Standardised testing and traditional grading systems may not fully capture the benefits of personalised and adaptive learning. Developing new evaluation frameworks that account for the diverse ways in which students learn and demonstrate knowledge is essential.

\subsection{Summary}

In this section, we reviewed the related work in the domains of education, social data analytics, and Generative AI. Our aim was to provide a contextual foundation for research on understanding and predicting the impact of Generative AI on the future of education. The areas covered include data curation in social media analytics, generative AI in education, impacts of AI on educational outcomes, shifts in educational and career choices, ethical and societal implications of AI in education, and future trends and predictions. We concluded with a summary comparing the methods and techniques reviewed.

Integrating AI into education impacts career choices, encouraging students to explore alternative educational pathways such as entrepreneurship. Personalised learning fosters autonomy and motivation, enabling students to develop practical skills through simulated real-world scenarios. However, ethical and societal concerns, including data privacy, algorithmic bias, and the digital divide, must be addressed to ensure equitable access to AI-driven educational resources. Looking ahead, the future of education in the age of Generative AI holds promising possibilities, with emerging trends like AI-driven tutoring systems, virtual and augmented reality, and predictive analytics playing significant roles in personalised and adaptive learning.

The next section will present the methodology used in this research, including social media data curation and deep-embedded clustering. This comprehensive approach aims to understand the acceptance and adoption of Generative AI in education, its impact on traditional education models, its influence on students' entrepreneurial ambitions, and educational outcomes associated with AI-driven personalised learning, providing valuable insights into the evolving educational landscape.


\section{Methodology}

This section outlines the methodology employed to understand and predict the impact of Generative Artificial Intelligence (AI) on the future of education. The methodology is designed in two key steps: contextualising social media data and introducing a novel approach to learning distributed representations and deep embedded clustering of texts to facilitate understanding the impact of Generative AI on the future of formal education.

\subsection{Contextualising Social Media Data}

The first step in our methodology involves leveraging ProcessGPT~\cite{beheshti2023processgpt} technology to contextualise social media data automatically. ProcessGPT, a state-of-the-art Generative Pre-trained Transformer (GPT)~\cite{yenduri2024gpt} model, is designed to generate human-like text and contextualise large volumes of raw data.
We fine-tuned this model specifically for the domain of education and Generative AI to ensure relevance and accuracy in the context of our research.

\subsubsection{Data Preprocessing}

In this phase, we preprocess the data by removing noise, irrelevant content, and duplicates. Preprocessing steps include~\cite{beheshti2017systematic} tokenisation, normalisation, and filtering based on relevance to education and Generative AI.
The initial step in preprocessing involves filtering out noise~\cite{beheshti2018crowdcorrect}. For example, in Twitter data, noise includes advertisements, spam tweets, and non-educational discussions. Tweets promoting unrelated products or services, such as "Check out this new phone case on sale!" are flagged as noise and discarded.
Next, we focus on removing irrelevant content. Using keywords and context analysis, we ensure that the data pertains specifically to Generative AI in education. For instance, a tweet discussing traditional teaching methods without mentioning AI, such as "Teachers should use more hands-on activities," is considered irrelevant and removed.

To handle duplicates, we employ algorithms~\cite{beheshti2017automating} that detect similar text patterns. For example, if the same tweet about a study on Generative AI in classrooms is reposted multiple times, only one instance is retained in the dataset.
After cleaning the data, we tokenise the text by breaking them down into individual words or tokens. For instance, the tweet "Generative AI enhances personalised learning experiences" is tokenised into ["Generative", "AI", "enhances", "personalised", "learning", "experiences"].
Normalisation is performed to standardise the text. This involves converting all characters to lowercase and removing punctuation. For example, "Generative AI!" and "generative ai" are both normalised to "generative ai".
Finally, we filter the data based on relevance using a relevance model. This model scores and retains content that specifically addresses the intersection of education and Generative AI. For example, a tweet titled "How Generative AI is Shaping the Future of Education" is kept, while a tweet about "AI in Healthcare" is filtered out.

By following these preprocessing steps, we ensure that the Twitter data is clean, relevant, and ready for further analysis or model training, focusing on the impact of Generative AI in the field of education.

\subsubsection{Contextualisation with ProcessGPT}

Once the data is preprocessed, we utilise ProcessGPT to transform this raw data into contextualised information. ProcessGPT employs techniques such as natural language processing (NLP) and deep learning to understand the context and generate meaningful representations of the data. The model is trained on a large corpus of educational content and fine-tuned to capture nuances specific to Generative AI in education. This step involves:
(i)~Feature Extraction: Identifying key features and entities within the social media data, such as topics, keywords, and sentiments,
(ii)~Enrichment: Linking extracted features to domain-specific knowledge bases, such as educational taxonomies and AI-related databases, to provide additional context, and
(iii)~Contextualisation: Creating enriched, contextualised data representations that are ready for further analysis.

\subsection{Learning Distributed Representations and Deep Embedded Clustering of Texts}

The second step of our methodology introduces a novel approach for Learning Distributed Representations and Deep Embedded Clustering of Texts~\cite{wang2023learning}. This approach facilitates a deeper understanding of the impact of Generative AI on student churn and the future of formal education.

\subsubsection{Distributed Representations}

To capture the semantic meaning of the contextualised social media data, we employ distributed representation techniques. These techniques transform textual data into continuous vector spaces, where semantically similar texts are positioned closer together. We utilise word embeddings such as Word2Vec, GloVe, and transformer-based embeddings like BERT~\cite{devlin2019bert} to generate these distributed representations. The process involves:
(i)~Embedding Generation: Converting text data into dense vector representations using pre-trained models, and
(ii)~Semantic Analysis: Analysing the embeddings to capture semantic similarities and differences among the data points.

\subsubsection{Deep Embedded Clustering}

To identify patterns and trends within the distributed representations, we apply Deep Embedded Clustering (DEC) techniques~\cite{guo2017improved}. DEC combines deep learning and clustering algorithms to perform end-to-end clustering of high-dimensional data. This involves:
Autoencoder Network: Using an autoencoder network to reduce the dimensionality of the data while preserving its important features;
Clustering Layer: Adding a clustering layer on top of the autoencoder to assign data points to different clusters based on their embeddings; and
Cluster Optimisation: Optimising the clustering process through iterative refinement, ensuring that the clusters are meaningful and representative of underlying patterns.

\subsubsection{Analysis of Student Churn and Educational Trends}

By clustering the contextualised social media data, we can identify distinct patterns related to student churn and educational trends influenced by Generative AI. This analysis includes:
(i)~Trend Detection: Identifying emerging trends and shifts in educational practices and career aspirations,
(ii)~Impact Assessment: Assessing the impact of Generative AI on student decisions to pursue entrepreneurial ventures over traditional university education, and
(iii)~Sentiment Analysis: Evaluating public sentiment and perceptions towards AI-driven education.


\subsection{Methodology}\label{Sec3}

The objective of this paper is to perform automatic text clustering. Specifically, given a set of diverse texts, the model should output the corresponding clustering results. To cluster similar assessments, BERT is employed to convert the texts into vector representations. To enhance deep feature extraction, a deep embedded clustering approach is utilized. To further improve clustering performance, Kullback-Leibler (KL) divergence~\cite{guo2017improved} and contrastive learning loss are incorporated during model training.

\subsubsection{Proposed Framework}

\begin{figure}[H]
\centering
\includegraphics[scale=0.5]{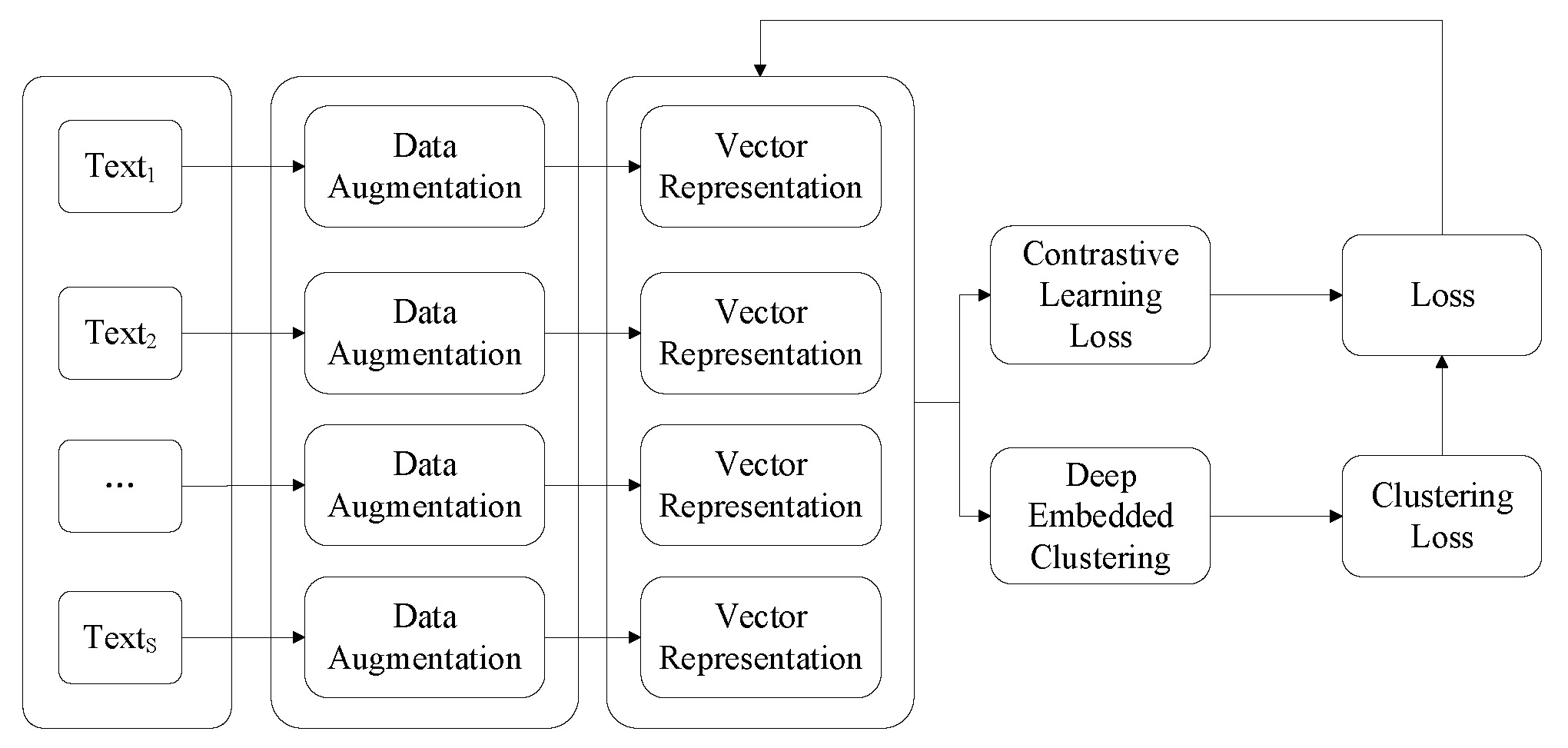}
\caption{\centering The proposed overall framework~\cite{wang2023learning}.} \label{fig1}
\end{figure}

To cluster similar texts, the overall framework of our proposed model is illustrated in Figure~\ref{fig1}. Initially, the texts are transformed into vectors using pre-trained language models (PLMs). These vectors are then clustered without specifying the number of clusters in advance. To enhance the representation quality, contrastive instances for the texts are generated by the vector representation models. For instance, consider an evaluation of assessments on the topic "Environmental Protection" at a middle school. Grading hundreds of these assessments is a substantial task for instructors. To address this challenge, the assessments are first augmented to increase their diversity. After the vector representation or clustering, the contrastive learning loss is calculated, which subsequently improves the representation performance.

\begin{figure}[H]
\centering
\includegraphics[scale=0.8]{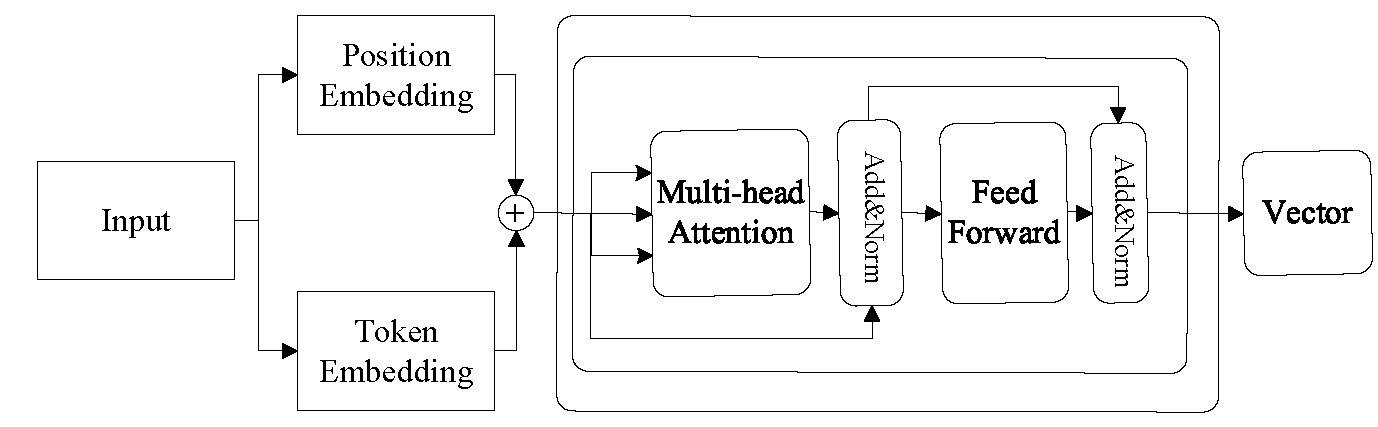}
\caption{\centering The procedure of text clustering~\cite{wang2023learning}.} \label{fig2}
\end{figure}

In this paper, we use the pre-trained BERT~\cite{devlin2018bert} to represent words and sentences in the given answers. We also explore the usage of the BiLSTM model~\cite{hochreiter1997long} for learning word-level and sentence-level information when we have few resources to fine-tune the whole BERT model.
The procedure of text clustering, including representation and clustering, is shown in Figure~\ref{fig2}.

\textbf{Vector Representation}
\vspace{2mm}

\noindent \textbf{BERT}.
In the input, BERT adds a special leading token \emph{[cls]} at the beginning of the input text (e.g., [cls] Artificial Intelligence is useful in education). \emph{[cls]} is used for sentence-level classification tasks during its pre-training stage, and its corresponding hidden vector is used as the sentence-level representations. All input tokens are represented as the additive combination of word embeddings and positional embeddings. Its core module is the multi-head self-attention~\cite{vaswani2017attention} where  input tokens are represented as queries ($Q_i=XW_i^Q$), keys ($K_i=XW_i^K$) and values ($V_i=XW_i^V$) in the $i^{th}$ head. We have:

\begin{align}
\mathit{SelfAttention}(Q_i,K_i,V_i)=softmax(\frac{Q_iK_i^T}{\sqrt{d_i^k }})V_i \\
\mathit{MultiHead}(Q,K,V)=\mathit{cat}(SelfAttention(Q_i,K_i,V_i))W^C
\end{align}

where $W_i^Q\in \Re^{d\times d_i^k}, W_i^K\in \Re^{d\times d_i^k},W_i^V\in \Re^{d\times d_i^v}$ $W^C\in \Re^{ \sum_{i=1}^{N_h}d_i^v\times d}$ are the trainable parameters. Finally, BERT has residual connection followed by layer-normalisation~\cite{ba2016layer}. A fully connected feed-forward network is applied to  matrix $Z_1$  as follows
\begin{equation}
  FF(Z_1)=max(0,Z_1W_1^F+b_1^F)W_2^F+b_2^F
\end{equation}

where  $W_1^F,b_1^F,W_2^F,b_2^F$ are parameter matrices. After Feed Forward, Add\&Norm is reused in the layer.

BERT performs well in transforming texts to vectors while it consumes a large amount of server resources. To solve the problem, DistilBERT is adopted in this paper which has the same general architecture as BERT~\cite{reimers2019sentence}. The token-type embeddings and the pooler are removed, while the number of layers
is reduced by a factor of 2. Most of the operations used in the Transformer architecture (linear
layer and layer normalisation) are highly optimised in modern linear algebra frameworks, and our investigations showed that variations on the last dimension of the tensor (hidden size dimension) have
a smaller impact on computation efficiency (for a fixed parameters budget) than variations on other
factors like the number of layers.

\subsubsection{Contrastive Learning}

Contrastive learning aims to maximise the similarities of
positive pair instances while it can maximise the difference of negative pair ones.
Various rules define the characteristics of pair instances. For example, when the pairs are within the same class, it is a positive pair. Otherwise, it is a negative pair. In
this paper, the positive and negative pairs are constructed at the
instance level by data
augmentations and embedding. In addition, the positive pairs consist
of samples either augmented or embedded from the same instance, while the
negative pairs are from different instances.

Considering a mini-batch $S$,  the instance contrastive learning loss is
defined in terms of the augmented pairs $S^{a}$ while for a mini-batch $S^{v}$, the instance contrastive learning loss is
defined in terms of the representation vectors.
$i\in\{1, \cdots, M\}$ is denoted as  the index of an arbitrary instance in  set $S$.
 $j\in \{1, \cdots , M\}$ is the index of the other instance in
$S^a$
augmented from the same instance in the original set $S$.
For pairs $(x_{i},x_{j}^{a})$, there are 2M pairs in total.
For a specific sample $x_{i} \in S$, there are 2M-1 pairs in total where $(x_{i},x_{i}^{a})$ is a positive pair and others are negative pairs.
Similarly, 
 $x_{j}^{v}\in \{1, \cdots , 2M\}$ is the embedding sample of the other instance in
$S^v$
embedded from the same instance in the original set $S$.
To alleviate the information loss induced by contrastive
loss, we do not directly conduct contrastive learning on
the feature matrix. Let $z_{i}$ and $z_{j}$ be the corresponding outputs of the
head $g$, i.e., $z_i = g(\varphi(x_i )), j = i_1
, i_2$.
Then for
$x_{i_1}$, we try to separate $x_{i_2}$ apart from all negative
instances in $S^a$ by minimising the following
applied. The pair-wise similarity is measured by cosine
distance:
\begin{equation}
    s(z_i, z_j)=\frac{z_i^Tz_j}{||z_i||_2||z_j||_2}
\end{equation}
To optimise pair-wise similarities, without loss of generality, the loss for a
given sample $x_i^a$
is described as follows {\color{red}~\cite{WOS:000681269800015}}:
\begin{equation}\label{equ_loss_co_i}
loss^{co}_i=-log\frac{exp(s(z_i, z_{i_1})\cdot\tau^{-1})}{\sum_{j=1}^{N}\texttt{1}_{j\neq i}\cdot exp(s(z_i, z_j)\cdot\tau^{-1})}
\end{equation}

where $tao$ is the temperature parameter that controls the difference's scale.
According to
Equation \eqref{equ_loss_co_i} and the mini-batch $S$, the contrastive loss is calculated using:
\begin{equation}\label{equ_ins_l}
   loss^{co}=\sum_{i=1}^{S}\frac{loss_i^{co}}{S}
\end{equation}

\subsubsection{Clustering}
To cluster similar instances together, there are different methods such as \emph{label as representation} and Kmeans algorithm in existing studies for deeply embedded clustering. \emph{Label as representation} is defined when projecting a data sample into a space where the dimensionality equals the number of clusters with the $i^{th}$ element interpreted as the probability of belonging to the
$i^{th}$ cluster, and the feature vector denotes its soft label accordingly. Kmeans algorithms have been used in ~\cite{miller2019leveraging,xie2016unsupervised,WOS:000681269800015} while the SOM algorithm with BERT has never been used in existing studies.

Self Organising Map ~\cite{kohonen1990self} is a classic unsupervised learning algorithm that follows the principle of topological mapping. The main aim of SOM is to convert
input signals of arbitrary dimension into one-dimensional or two-dimensional discrete mapping and the spatial position of the output layer neurons in the topological
mapping corresponds to a specific feature of input space.
The original idea of SOM is to simulate how vision systems work in the human brain, which is used for the organisation and visualisation of complex data. In general,
SOM can capture the topology map structure which distinguishes the distances among different clusters and distribution of the
input data to provide a clustering analysis.
In this paper, a $M\times N$ self-organising map is constructed. The architecture
of SOM includes two layers: the input layer and the Kohonen output layer.

To describe SOM clearly,
$x$ is denoted as the input vector while  $W_{ij}(i\in\{1,\cdots,M\}, j\in\{1,\cdots,N\})$ is one of the neurons in the self-organisation map which is a unit vector.
In addition, the shape of weights neurons is the same as the input vector $z_k$.
To calculate the distance from $z_k$ to the weights neurons, $z_k$ is reduced to as a unit vector where $z_{kj}$ is described as follows:
\begin{equation}
    z_{kj}=\frac{z_{kj}}{\sum_{j=1}^{d}z_{kj}}
    \label{equ_7}
\end{equation}
For each sample $z_k$, the distance $d_{ij}$ from  a neurons $W_{ij}$ is calculated by:
\begin{equation}
    d_{ij}=||z_k-W_{ij}||^2
\end{equation}
The corresponding neuron which has the minimum distance is called the winner.
The rates of the modifications at different nodes depending on the
mathematical form of the function $h_{ci}(t)$ which is described as follows~\cite{kohonen2013essentials}:
\begin{equation}
    h_{ci}(t)=a(t)exp(\frac{-sqdist(c,i)}{2\delta (t)})
    \label{equ_9}
\end{equation}
where $\alpha(t)$ is the learning rate at time $t$, $-sqdist(c,i)$ is the square of the geometric distance between the nodes $c$ and $i$ in the grid, and $\delta (t)$ is a monotonically decreasing function of $t$ which controls the number of neighborhood nodes.

\subsubsection{Clustering Loss}
 Suppose our data consists of $M\times N$
semantic categories and each category is characterised by its centroid in the $M\times N$ representation space,
denoted as $\mu_{ij}~ (i\in\{1,2,\cdots,M\},j\in\{1,2,\cdots,N\})$. $e_j=g(x_j)$ is denoted as the representation of the instance $x_j$ in the original set $S^v$.
According to~\cite{van2008visualizing}, the student's t-distribution is used to compute the
probability $q_{ij}$ of assigning $x_j$ to one of the neurons $W_{ij}$:
\begin{equation}
    q_{ji}=\frac{(1+\frac{||e_j-\mu_i||_2^2}{\alpha})^{-\frac{\alpha+1}{2}}}{\sum_{k=1}^{M\times N}(1+\frac{||e_j-\mu_k||_2^2}{\alpha})^{-\frac{\alpha+1}{2}}}
    \label{equ_10}
\end{equation}
where $\alpha$ denotes the degree of freedom of the Student's t-distribution and $\alpha=1$ by default~\cite{van2008visualizing}.
According to SOM, the neurons are obtained. The neurons are refined by leveraging an
auxiliary distribution proposed by~\cite{xie2016unsupervised}.
$p_{ji}$ is denoted as the auxiliary distribution, which is described as follows:
\begin{equation}
  p_{ji}=\frac{\frac{q_{ji}}{f_i}}{\sum_{i^{'}}\frac{q_{ji^{'}}  }{f_{i{'}}}}
\end{equation}
where $f_{i}=\sum_{i=1}^{M}q_{jk}, k\in\{1,\cdots,K\}$ is the  interpreted as the soft cluster frequencies approximated
within a minibatch.
The target distribution first
sharpens.
The soft-assignment probability $q_{ji}$ is sharpened  by
raising it to the second power, and it is normalised
by the associated cluster frequency. Learning is encouraged between high confidence
cluster assignments
and  the bias which is simultaneously combated by imbalanced clusters.

To optimise the KL
divergence between
the cluster assignment probability and the target distribution, the cluster loss for each instance is described as follows:
\begin{equation}\label{equ_loss_cl}
    loss_i^{c}=KL[p_j||q_j]=\sum_{i=1}^{K}p_{ji}log\frac{p_{ji}}{q_{ji}}
\end{equation}

Since the mini-batch instances are studied in this paper, the cluster loss for each mini-batch is obtained as follows in terms of Equation \eqref{equ_loss_cl}:

\begin{equation}\label{equ_loss_c_all}
    loss^{c}=\sum_{i=1}^{M}\frac{loss_i^{c}}{M}
\end{equation}

The overall loss function is obtained by Equations \eqref{equ_ins_l} and  \eqref{equ_loss_c_all} which is described as follows:
\begin{equation}
    l=loss^{co}+loss^{c}
\end{equation}

\subsection{Algorithms}

Distill BERT is first used to transform texts into vectors with a text augmentation technique to cluster similar assessments.
After augmentation, contrastive learning and clustering techniques are adopted to cluster similar texts.
Since the efficiency of different clustering algorithms is different,
we propose three different clustering algorithms with the proposed model.
The proposed algorithm framework is described in Algorithm \ref{algorithm_1}.
To calculate the efficiency of Algorithm \ref{algorithm_1}, the adopt Accuracy (ACC)
and Normalised Mutual Information (NMI) are measured.
NMI is calculated by NMI$(Y,C)=\frac{2\times I(Y;C)}{H(Y)+H(C)}$ where $Y$ is the real label, $C$ is the clustering label, H(.) is the cross enropy, and I(.;.) is the mutual information.

To describe the algorithm framework clearly in Algorithm~\ref{algorithm_1},
the input data contains dataset $S$, training epochs $E$, temperature $\tau$ in contrastive learning and the structure of augmentation while the output results are ACC and NMI.
A mini-batch $\{x_i\}$ of texts is selected from dataset $S$. For each input sample $x_i$,  two random augmentations are generated where $x_i^a = A^S_j (x_i)(j=1,2)$,  which represents a different view of the text and contains some
subset of the information in the original sample. $x_i^1,x_i^2$ ($i\in\{1,2,\cdots,|S|\}$) are denoted as
the augmented text of an assessment $x_i$.
After text augmentation, the contrastive loss is calculated according to the temperature $\tau$. Texts are transformed into vectors by DistillBERT.
The predicted labels are obtained after the clustering algorithms.
Compared to predicted labels to ground truth labels, the clustering loss is calculated.
According to the instance loss and clustering loss, the overall loss is obtained.
The ACC and  NMI metrics are calculated.

\begin{algorithm}[!tb]
\begin{small}
\KwIn{Data set $S$, Training epochs $E$, Temperature $tau$, Structure of augmentation $A^S$}
 \KwOut{ ACC, NMI.}
 \For{$e=1$ to $E$ }{
 Sample a mini-batch \{$x_i$\}($i\in\{1,2,\cdots,M\}$) from dataset $S$;\\
 Sample two augmentations $A^S_1, A^S_2$;\\
 \For{$i=1$ to $|S|$ }{
 $x_i,x_i^1,x_i^2$ is represented by DistilBERT;\\
 }
 Compute the instance loss by Equation \eqref{equ_loss_co_i};\\
 Calculate the clustering centres;\\
 Cluster these vectors by different clustering methods;\\
 Calculate the predicted labels for all vectors;\\
 Compute the clustering loss by Equation \eqref{equ_loss_cl};\\
 Compute the overall loss by Equation \eqref{equ_loss_c_all};\\
 Calculate the
ACC, NMI by comparing predicted labels to ground truth labels;\\
 Update gradients
 to minimise the overall loss;\\}
 \Return ACC, NMI.
 \caption{Deep Embedded Clustering Framework}
 \label{algorithm_1}
\end{small}
\end{algorithm}

The deeply embedded clustering framework is described in Algorithm \ref{algorithm_1}.
For vector clustering techniques, there are three different strategies: label as representation, Kmeans algorithm, and SOM algorithm.
For label as representation, the dimension of vectors is projected to the number of clusters by multiple neuron networks.
The index of the maximum values of the vector is represented as the label of clusters.
After all of the vectors are projected to the fixed dimensions, Kmeans and SOM algorithms are used to cluster similar vectors together, respectively.
To calculate the clustering centres for the label as representation algorithm, Kmeans and SOM algorithms are adopted respectively.

\begin{algorithm}[!tb]
\begin{small}
\KwIn{Matrices $Z,Z^1,Z^2$}
  Initialise labels $L,L_1, L_2$;\\
 \For{$i=1$ to $|S|$ }{
 $z_i$ is projected to $K$ dimensions;\\
 $z_i^1$ is projected to $K$ dimensions;\\
 $z_i^2$ is projected to $K$ dimensions;\\
 $ind_{max}\leftarrow 1$, $ind_{max}^1\leftarrow 1$, $ind_{max}^2\leftarrow 1$;\\
  $z_{max}\leftarrow z_i[1]$, $z_{max}^1\leftarrow z_i^1[1]$, $z_{max}^2\leftarrow z_i^2[1]$;\\
  \For{$j=2$ to $K$ }{
  \If{$z_{max}<z_i[j]$}
  {$z_{max}\leftarrow z_i[j]$;\\
  $ind_{max}\leftarrow j$;
   }
   }
   $L_i\leftarrow ind_{max}$;\\
    \For{$j=2$ to $K$ }{
  \If{$z_{max}^1<z_i^1[j]$}
  {$z_{max}^1\leftarrow z_i^1[j]$;\\
  $ind_{max}^1\leftarrow j$;
   }
   }
    $L_i^1\leftarrow ind_{max}^1$;\\
    \For{$j=2$ to $K$ }{
  \If{$z_{max}^2<z_i^2[j]$}
  {$z_{max}^2\leftarrow z_i^2[j]$;\\
  $ind_{max}^2\leftarrow j$;
   }
   }
   $L_i^2\leftarrow ind_{max}^2$;\\
  }
 \Return $L,L_1,L_2$.
 \caption{Label as Representation Algorithm}
 \label{algorithm_2}
\end{small}
\end{algorithm}


\begin{algorithm}[!bt]
\begin{small}
\KwIn {Matrices $Z,K$}
Initialise  $K$ centroids $c_1,\cdots,c_K$ randomly;\\
$t \leftarrow 0$;\\
\While{Conditions are satisfied}{
$c_1'\leftarrow c_1,\cdots,c_K'\leftarrow c_K$;\\
Initialise  empty sets $C_1,\cdots,C_K$;\\
\For{$i=1$ to  $|S|$}{
$d_1 \leftarrow ||z_i-c_1||^2$, $ind\leftarrow 1$;\\
\For{$j=2$ to $K$}{
 \If{$d_1<||z_i-c_j||^2$}
 {$d_1\leftarrow ||z_i-c_j||^2$;\\
 $ind\leftarrow j$;\\}
}
$C_j\leftarrow C_j\cup z_{ind}$;\\
}
$c_1\leftarrow avg(C_1),\cdots,c_K\leftarrow avg(C_K)$;\\
$t\leftarrow t+1$
}
 \Return $L$.
 \caption{Kmeans Clustering Algorithm}
 \label{algorithm_3}
\end{small}
\end{algorithm}

\begin{algorithm}[!bt]
\begin{small}
 \KwIn {$M, N, Z$, iteration times $T$,$\alpha$}
 Initialise the weights  $W(0)$;\\
 Normalise $Z$ by equation \eqref{equ_7};\\
 \For{$t=1$ to $T$}{
 \For{$i=1$ to $|S|$}{
 $d\leftarrow 100$, $L\leftarrow 0_{1\times|S|}$;\\
 \For{$j=1$ to $M$}{
 \For{$k=1$ to $N$}{
 \If{$||z_i-W_{jk}||^2<d$}{$d\leftarrow||z_i-W_{jk}||^2$;\\}
 $L_i\leftarrow (j-1)*N+k$;\\
 }
 }
 }
 Calculate $h_{ci}(t)$ by Equation \eqref{equ_9};\\
 Update learning rate $l_r(t)=\alpha *(1-\frac{t}{T})$;\\
 Update weights by Equation \eqref{equ_10};\\

 }

 \Return $L$.
 \caption{SOM Clustering Algorithm}
 \label{algorithm_4}
 \end{small}
\end{algorithm}

The label as representation algorithm is shown in Algorithm \ref{algorithm_2}. The input of the algorithm is the matrices $Z, Z_1$, and $Z_2$ which are made of $z_i,z_i^1,z_i^2(i\in\{1,\cdots,|S|\})$ where $z_i,z_i^1,z_i^2$ are represented by DistilBERT.
The labels $L,L_1,L_2$ are initialised first.
According to the idea of the label as representation, all the represented vectors are projected to $K$ dimensions. The index of the maximum value has selected the label for the projected vectors. The time complexity of the Algorithm \ref{algorithm_2} is $O(K|S|)$.

 The Kmeans clustering algorithm is described in Algorithm \ref{algorithm_3}. To describe Algorithm \ref{algorithm_3} clearly, $c_1,\cdots,c_k$ are  denoted as the initial centroids.
 All the vectors in $Z$ are compared with $K$ centroids. The vector with the minimum distance to the centroid $c_i$ is assigned to $C_i$.
 After all the vectors are assigned, new centroids are calculated in terms of new $C_i$ ($i\in\{1,\cdots,K$).
 The procedure is repeated until the conditions are not satisfied. The label $L$ for $Z$ is obtained with the final centroids.
 The labels $L_1, L_2$ for $Z_1$ and $Z_2$ are obtained, similarly.
 The time complexity for Algorithm \ref{algorithm_3} is $O(|S|Kt)$.

The SOM clustering algorithm is described in Algorithm \ref{algorithm_4}.
 The input data includes parameters $M,N$ for the map, the input matrix $Z$, the iteration times $T$, and the parameter $\alpha$.
 The weight matrix $W$ is initialised where each raw is a unit vector.
 To compare the input matrix $W$ to $Z$, each raw of $Z$ is normalised to a unit vector.
By comparing $Z$ to $W$, the distance from $z_i$ to all the neurons is calculated.
The neuron with the minimum distance is selected as the winner.
The label of $z_i$ is determined.
The learning rate is updated.
After that, $h_{ci}(t)$ is calculated by Equation \eqref{equ_9} with new $\delta(t)$.
The weights are updated by the new $h_{ci}(t)$.
The procedure is repeated until $t=T$.
The time complexity of Algorithm \ref{algorithm_4} is $O(MN|S|T)$.


\section{Experiments and Evaluation}

In this section, the experimental
setup
is described
and different clustering strategies are compared.

\subsection{Experimental Setup}

In the proposed deep-embedded clustering algorithm, there are different parameters and clustering components and we will  present our experiments to calibrate the parameters and components using
different datasets which include the AgNews dataset ~\cite{zhang2015text},
 and StackOverflow dataset \cite{XU201722}. 
The AgNews dataset is a subset of news titles that  contains 4 topics. AgNews is a collection of more than 1 million news articles which  have been gathered from more than 2000 news sources. The dataset is used in data mining including clustering, classification, information retrieval and so on.
StackOverflow is a subset of the challenge
data published by
Kaggle, which contains 20,000
question titles associated with 20 different categories~\cite{XU201722}.  For each question, it includes: Question ID, Creation date,
Closed date,
Number of answers and so on.
We then present experimental results to compare the proposed deep-embedded clustering algorithm
against three existing algorithms
using real-life instances.
Different algorithms are implemented  in the Python programming language with the Pytorch library on a single NVIDIA P100 GPU with 16G memory.

The Adam optimiser
with a batch size of 400,  the distilbert-basenli-stsb-mean-tokens, and the  maximum input length are used the same as in \cite{zhang2021supporting}. The distilbert-basenli-stsb-mean-tokens in the Sentence Transformers
library is used   as the backbone~\cite{reimers2019sentence}, and we set the maximum input length to
32. Different learning rates are conducted to optimise both the Clustering head and
we set $\alpha=1$ for the dataset.
As mentioned in Section 3, different $tau$ are used to optimise
the contrastive loss. We tried different $tau$ values in
the range of (0, 1]. For fair
comparison between deeply embedded clustering framework and its components or
variants, the clustering performance is analysed for
each of them by applying Kmeans, label as a representation with clustering centres by
Kmeans (KmeansR), and  SOM (SOMR), are labelled as representations with clustering centres by SOM, and SOM respectively.

\subsection{Algorithm Comparison}

To analyse the experimental results on the Agnews  and StackOverflow datasets, different parameters such as learning rates, learning rate scale (lr scale), and the temperature $tau$ are calibrated  and compared with different values on these different algorithm components.

\subsubsection{Experimental Results on Learning Rates}

\begin{table*}[!tb]
\begin{small}
 \caption{Learning rate  Calibration and Algorithm Component Comparison ($10^{-7}-10^{-3}$).}
 \centering
 \renewcommand{\tabcolsep}{1.5pt}
  \label{T1}
 \begin{tabular}{p{2.2cm}p{1.5cm}p{1.7cm}p{1.2cm}p{1.2cm}p{1.2cm}p{1.2cm}p{1.2cm}p{1.2cm}}
 \hline
  \toprule
Dataset&Metrics&Algorithm&$10^{-7}$&$10^{-6}$&$10^{-5}$&$10^{-4}$&$10^{-3}$\\
&NMI &SOM&0.763&0.739 &\textbf{0.824}&0.565&0.625\\
&NMI&SOMR&0.724&0.729&0.696&0.491&0.743\\
&NMI &Kmeans&0.767&0.744 &0.822&0.731&0.454\\
&NMI&KmeansR&0.727&0.731&0.685&0.718&0.296\\
Avgnews&ACC &SOM&0.922&0.909 &\textbf{0.947}&0.804&0.836\\
&ACC&SOMR&0.899&0.903&0.887&0.767&0.914\\
&ACC &Kmeans&0.923&0.926 &0.944&0.899&0.694\\
&ACC &KmeansR&0.901&0.904 &0.881&0.865&0.494\\
\bottomrule
 \end{tabular}
 \end{small}
\end{table*}

To calibrate the learning rates, the values are set from $10^{-7}$ to $10^{-3}$ for the Avgnews dataset.
The experimental results  are shown in Table \ref{T1}.
According to Table \ref{T1}, With NMI metrics, when the learning rate is $10^{-5}$, the NMI is the highest with 0.824 while for SOMR, the highest NMI is 0.743 with learning rate $10^{-3}$.
For K-means algorithm, the highest NMI is 0.822 with a learning rate $10^{-5}$ while for KmeansR, the highest NMI is 0.731 with a learning rate $10^{-6}$.
Comparing SOM, SOMR, Kmeans, and KmeansR, the highest NMI is 0.824 obtained by SOM, followed by 0.822 obtained by Kmeans.
For ACC, the highest value of SOM is 0.947 with the learning rate $10^{-5}$ while it is 0.915 of SOMR with the learning rate $10^{-3}$. For Kmeans, the highest value of ACC is 0.944 with the same learning rate $10^{-5}$ as SOM while for SOMR, the highest ACC is 0.904 with the learning rate $10^{-6}$.
Comparing all algorithm components, the highest value is 0.947 obtained by SOM.

\begin{figure}[!tb]
\centering
\includegraphics[scale=0.25]{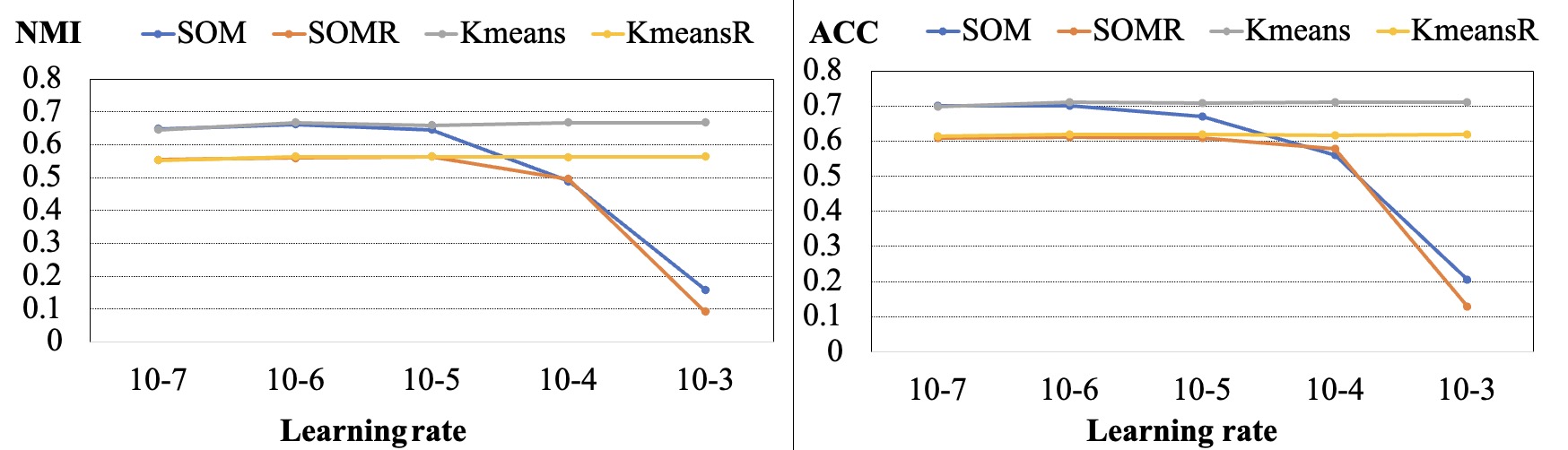}
\caption{Algorithm Componenet Comparison with StackOverflow for NMI and ACC.} \label{fig3}
\end{figure}

To calibrate the learning rates for the StackOverflow dataset,
the experimental results  are shown in Figure \ref{fig3}.
According to Figure \ref{fig3}, the SOM and SOMR algorithms are affected by the learning rates while the Kmeans and KmeansR algorithm has a little effect on the  learning rates. With NMI metrics, when the learning rate is $10^{-6}$, the NMI is the highest with 0.662 while for SOMR, the highest NMI is 0.564 with learning rate $10^{-5}$.
For Kmeans algorithm, the highest NMI is 0.667 with learning rates $10^{-6}$, $10^{-4}$, and $10^{-3}$ while for KmeansR, the highest NMI is 0.564 with the same corresponding learning rates.
Comparing SOM, SOMR, Kmeans, and KmeansR, the highest NMI is 0.667 obtained by Kmeans, followed by 0.662 obtained by SOM.
For ACC, the highest value of SOM is 0.710 with the learning rate $10^{-6}$ while it is 0.611 of SOMR with the learning rate $10^{-6}$. For Kmeans, the highest value of ACC is 0.711 with the same learning rate $10^{-6}$ as SOM while for SOMR, the highest ACC is 0.619 with the learning rate $10^{-6}$, $10^{-5}$, and $10^{-3}$.
Comparing all algorithm components, the highest value is 0.711 obtained by the Kmeans algorithm.

\begin{table*}[!tb]
\begin{small}
 \caption{Learning Rate Scale Calibration and Algorithm Component Comparison (50-500) .}
 \centering
 \renewcommand{\tabcolsep}{0.2pt}
  \label{T2}
 \begin{tabular}{p{2.2cm}p{2.0cm}p{1.2cm}p{1.2cm}p{1.2cm}p{1.2cm}p{1.2cm}p{1.2cm}p{1.2cm}p{1.2cm}p{1.2cm}}
 \hline
  \toprule
Dataset&Algorithms&Metrics&50&100&150&200&250&500\\
&SOM&NMI &0.711&0.824 &0.827&0.829&0.862&0.883\\
&SOMR&NMI&0.609&0.694&0.670&0.662&0.662&0.712\\
AvgNews&Kmeans&NMI &0.823&0.822 &0.836&0.844&0.856&\textbf{0.904}\\
&KmeansR&NMI&0.699&0.685&0.664&0.663&0.669&0.744\\
&SOM&ACC &0.901 &0.947&0.947&0.947&0.961&0.968\\
&SOMR&ACC&0.882&0.837&0.873&0.871&0.871&0.891\\
&Kmeans&ACC &0.944&0.944 &0.950&0.953&0.958&\textbf{0.974}\\
&KmeansR&ACC &0.888&0.881 &0.870&0.870&0.868&0.905\\
  \bottomrule
 \end{tabular}
 \end{small}
\end{table*}

\subsubsection{Experimental Results on Learning Rate Scales}

To calibrate the learning rate scale on different clustering algorithms for the Avgnews dataset, the values are set from $50$ to $500$.
The experimental results  are shown in Table \ref{T2}.
According to Table \ref{T2}, with NMI metrics, when the learning rate scale is 500, the NMI for the Avgnews dataset is the highest with 0.883 for SOM and for SOMR, the highest NMI is 0.712.
For the Kmeans algorithm, the highest NMI is 0.904 with a learning rate scale 500 and for KmeansR, the highest NMI is 0.744 with the same learning rate scale.
Comparing SOM, SOMR, Kmeans, and KmeansR, the highest NMI is 0.904 obtained by $Kmeans$, followed by 0.883 obtained by SOM.
According to Table \ref{T2}, with ACC metrics, when the learning rate scale is 500, the NMI is the highest with 0.968 for SOM and for SOMR, the highest NMI is 0.891.
For the K-means algorithm, the highest ACC is 0.974 with a learning rate scale 500 and for KmeansR, the highest ACC is 0.905 with the same learning rate scale.
Comparing SOM, SOMR, Kmeans, and KmeansR, the highest NMI is 0.974 obtained by Kmeans, followed by 0.968 obtained by SOM.

\begin{figure}[H]
\centering
\includegraphics[scale=0.25]{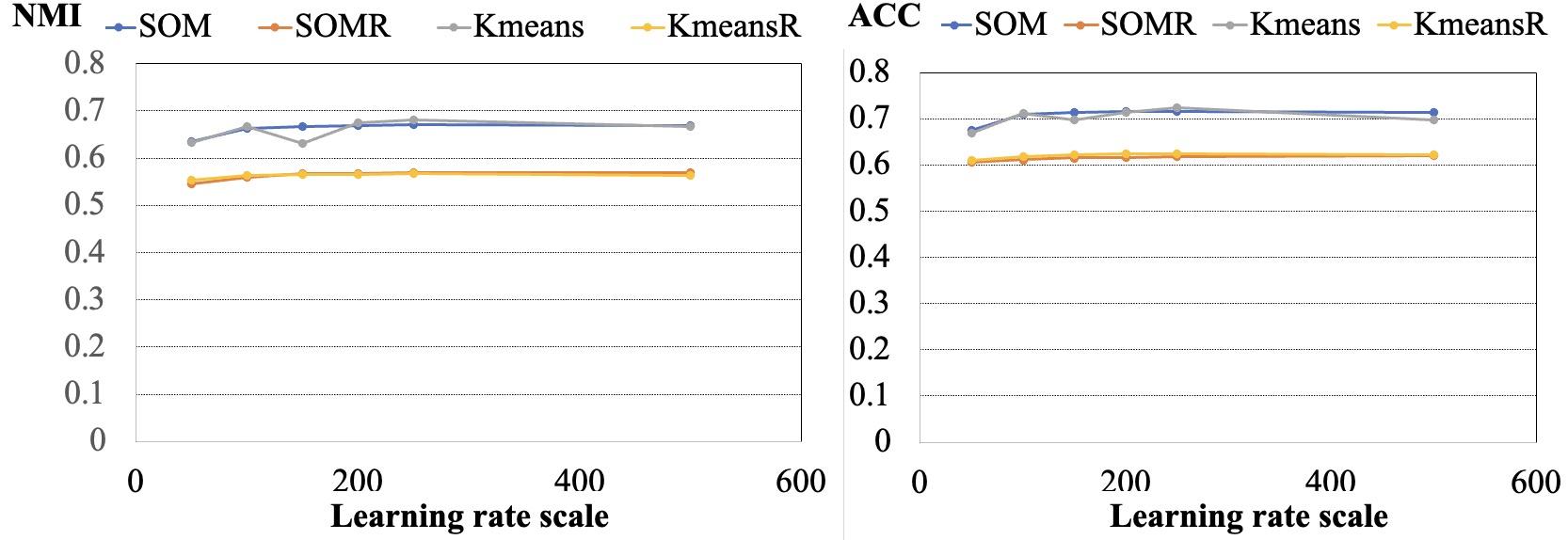}
\caption{Learning Rate Scale Calibration and Algorithm Componenet Comparison with StackOverflow for NMI and ACC
.} \label{fig4}
\end{figure}

According to Figure \ref{fig4}, the SOM algorithm is a little better than Kmeans algorithm in NMI and ACC metrics. Similarly, SOMR is slightly better than KmeansR. For the StackOverflow dataset, with NMI metrics, when the learning rate scale is 250, the NMI for the StackOverflow dataset is the highest with 0.670 for SOM and for SOMR, the highest NMI is 0.570.
For the K-means algorithm, the highest NMI is 0.680 with a learning rate scale 250, and for KmeansR, the highest NMI is 0.568 with the same learning rate scale.
Comparing SOM, SOMR, Kmeans, and KmeansR, the highest NMI is 0.680 obtained by Kmeans, followed by 0.670 obtained by SOM.
According to Figure \ref{fig4}, with ACC metrics, when the learning rate scale is 250, the NMI is the highest with 0.716 for SOM and for SOMR, the highest NMI is 0.620 with a learning rate scale of 500.
For the K-means algorithm, the highest ACC is 0.724 with a learning rate scale of 500 and for KmeansR, the highest ACC is 0.624 with the same learning rate scale.
Comparing SOM, SOMR, Kmeans, and KmeansR, the highest NMI is 0.724 obtained by Kmeans, followed by 0.716 obtained by SOM.

\subsubsection{Experimental Results on Temperature}
\begin{table*}[!tb]
\begin{small}
 \caption{Temperature  Calibration and Algorithm Component Comparison. (0.4-0.9)}
 \centering
 \renewcommand{\tabcolsep}{0.2pt}
  \label{T3}
 \begin{tabular}{p{2.2cm}p{2cm}p{1.2cm}p{1.2cm}p{1.2cm}p{1.2cm}p{1.2cm}p{1.2cm}p{1.2cm}}
 \hline
  \toprule
Dataset&Temperature&Metrics&0.4&0.5&0.6&0.7&0.8&0.9\\
&SOM&NMI &0882&0.883 &0.881&0.948&0.958&\textbf{0.966}\\
&SOMR&NMI&0.679&0.712&0.735&0.784&0.821&0.844\\
&Kmeans&NMI &0.881&0.904 &0.905&0.905&0.909&0.953\\
Avgnews&KmeansR&NMI&0.765&0.744&0.764&0.784&0.807&0.819\\
&SOM&ACC &0.967&0.968 &0.967&0.987&0.990&\textbf{0.992}\\
&SOMR&ACC&0.874&0.891&0.902&0.920&0.938&0.949\\
&Kmeans&ACC &0.966&0.974 &0.975&0.975&0.976&0.989\\
&Kmeans&ACCR &0.885&0.905 &0.914&0.922&0.931&0.939\\
  \bottomrule
 \end{tabular}
 \end{small}
\end{table*}

To calibrate the temperature $tau$, the values are set from $0.4$ to $0.9$.
The experimental results are shown in Table \ref{T3}.
According to Table \ref{T3}, with NMI metrics, when the temperature is 0.9, the NMI is the highest with 0.966 for SOM and for SOMR, the highest NMI is 0.821.
For the K-means algorithm, the highest NMI is 0.909 with a temperature of 0.9 and for KmeansR, the highest NMI is 0.907 with the same temperature.
Comparing SOM, SOMR, Kmeans, and KmeansR, the highest NMI is 0.966 obtained by Kmeans, followed by 0.953 obtained by Kmeans.
According to Table \ref{T3}, with ACC metrics, when the temperature is 0.9, the ACC is the highest with 0.992 for SOM and for SOMR, the highest NMI is 0.949.
For the K-means algorithm, the highest ACC is 0.989 with a temperature of 0.9 and for KmeansR, the highest ACC is 0.939 with the same temperature.
Comparing SOM, SOMR, Kmeans, and KmeansR, the highest ACC is 0.992 obtained by SOM, followed by 0.989 obtained by Kmeans.

\begin{figure}[!bt]
\centering
\includegraphics[scale=0.25]{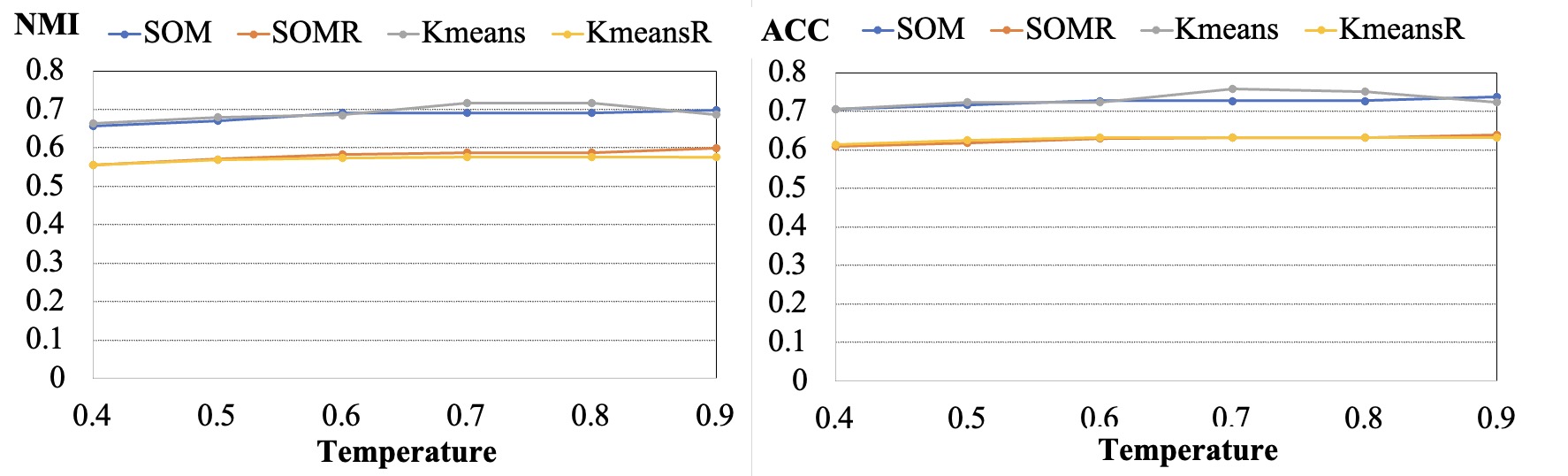}
\caption{Learning Rate Scale Calibration and Algorithm Componenet Comparison with StackOverflow for NMI and ACC
.} \label{fig5}
\end{figure}

The experimental results  are shown in Figure \ref{fig5} for the StackOverflow dataset.
According to Figure \ref{fig5}, parameter temperature has a little effect on SOM, SOMR, Kmeans, and KmeansR algorihtms. With NMI metrics, when the temperature is 0.9, the NMI is the highest with 0.699 for SOM and for SOMR, the highest NMI is 0.598.
For the K-means algorithm, the highest NMI is 0.716 with a temperature of 0.7 and for KmeansR, the highest NMI is 0.577 with the same temperature.
Comparing SOM, SOMR, Kmeans, and KmeansR, the highest NMI is 0.716 obtained by Kmeans, followed by 0.699 obtained by SOM.
According to Figure \ref{fig5}, with ACC metrics, when the temperature is 0.9, the ACC is the highest with 0.737 for SOM and for SOMR, the highest NMI is 0.639.
For Kmeans algorithm, the highest ACC is 0.757 with a temperature of 0.7 and for KmeansR, the highest ACC is 0.632 with the same temperature.
Comparing SOM, SOMR, Kmeans, and KmeansR, the highest ACC is 0.757 obtained by Kmeans, followed by 0.737 obtained by SOM.

According to above experiments, Kmeans and SOM algorithms are always better than SOMR and KmeansR.
For the Avgnews dataset, SOM performs the best while for the StackOverflow dataset, Kmeans performs the best.

\subsection{Evaluation of Twitter Dataset}

In this section, we present the evaluation of our proposed approach using a Twitter dataset.
This dataset contains tweets that was gathered by crawling a REST API using the Python library tweepy\footnote{https://www.tweepy.org/}, a Python library designed to simplify interaction with the Twitter API (now known as X\footnote{https://x.com/}). Tweepy provides a straightforward and efficient way for developers to programmatically access and interact with Twitter data. With Tweepy, tasks such as fetching tweets, posting updates, managing Twitter accounts, and streaming live data become significantly easier. The library handles authentication, API requests, and even streaming real-time tweets, making it popular among data scientists, developers, and researchers working with social media data. Common use cases include sentiment analysis, tracking hashtags, automating posts, and monitoring trends. By offering a clean and intuitive interface, Tweepy allows users to focus on building their applications or conducting analysis without getting bogged down by the complexities of API interactions.

We analyse the effectiveness of our methodology in contextualising tweets related to education and extracting insights on the impact of Generative AI on education. The evaluation involves measuring the accuracy of feature extraction, sentiment analysis, and trend detection. We also highlight the important features used to understand the relevance of tweets to the education domain.
The dataset consists of 50,000 tweets collected over a six-month period containing keywords related to education and Generative AI. The tweets were pre-processed to remove duplicates, irrelevant content, and noise. The resulting dataset was used for both training and evaluation purposes.
Table~\ref{tblFeature} illustrates the important features and insights extracted from the tweets.

\begin{table}
 \caption{The important features and insights extracted from the tweets.}
 \centering
  \begin{tabular}{cc}
   \includegraphics[scale=0.8]{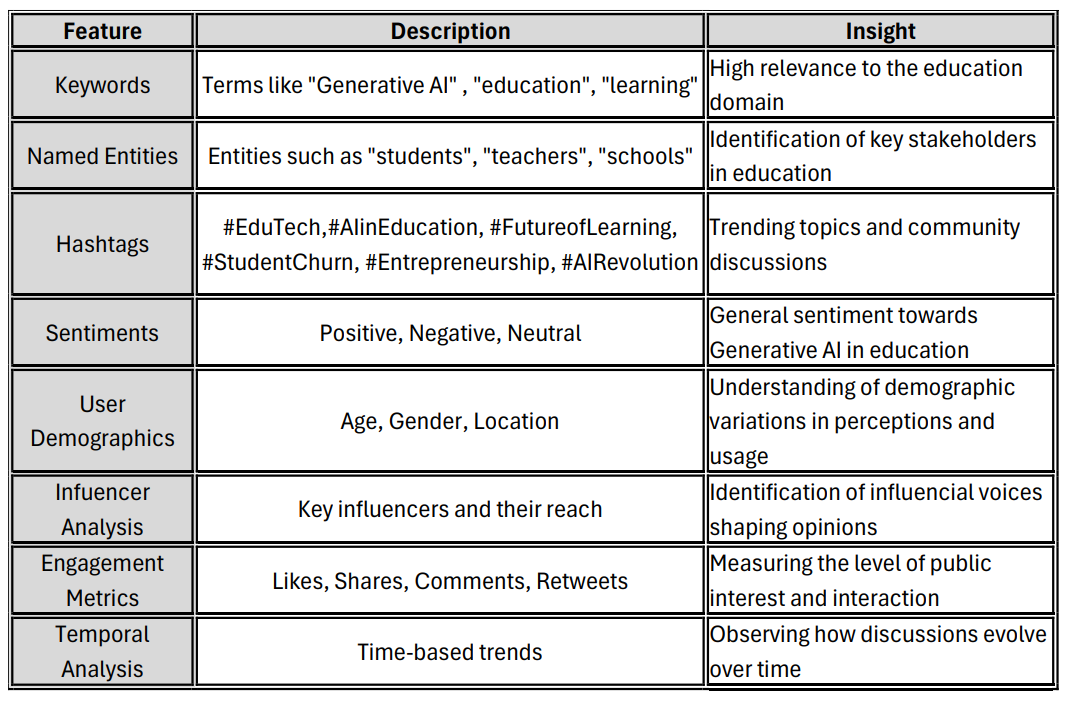}\\
  \end{tabular}
 \label{tblFeature}
\end{table}

\textbf{Keywords.}
The keywords extracted from the tweets, such as "Generative AI," "education," and "learning," were crucial in determining the relevance of the content to the education domain. These keywords helped in filtering out non-relevant tweets and focusing on discussions pertinent to our study. The high precision and recall rates indicate that our model effectively identified tweets containing these key terms, providing a solid foundation for further analysis.

\textbf{Named Entities.}
Identifying named entities like "students," "teachers," and "schools" was essential for understanding the context and stakeholders involved in the discussions about Generative AI in education. These entities allowed us to categorise tweets based on who was being talked about or who was the subject of the tweet, thereby helping to segment the data for more targeted analysis. This segmentation is vital for deriving insights specific to different groups within the education ecosystem.

\textbf{Hashtags.}
Analysing hashtags such as \#EduTech, \#AIinEducation, \#FutureofLearning, \#StudentChurn, \#Entrepreneurship, and \#AIRevolution provided insights into the trending topics and the broader community discussions. Hashtags are powerful indicators of the themes and issues that resonate with the public. They help in tracking the spread and popularity of specific topics, which is crucial for understanding the reach and impact of discussions related to Generative AI in education.

\textbf{Sentiments.}
Sentiment analysis was used to classify the emotional tone of the tweets as positive, negative, or neutral. Understanding the sentiment helped gauge the general perception of Generative AI in education. For instance, positive sentiments often highlighted the benefits and excitement about AI-driven personalised learning, while negative sentiments raised concerns about data privacy and algorithmic bias. Neutral sentiments provided a balanced view, focusing on factual information and observations.

\textbf{User Demographics.}
Analysing user demographics, including age, gender, and location, provided insights into the variations in perceptions and usage of Generative AI across different demographic groups. For example, younger users might be more enthusiastic about adopting AI technologies in education, while older users might be more cautious or skeptical. Geographic analysis helped identify regional trends and differences in how Generative AI is perceived and implemented in education systems around the world.

\textbf{Influencer Analysis.}
Identifying key influencers and their reach helped understand who drives the conversation about Generative AI in education. Influencers with significant followings can shape public opinion and drive engagement. By analysing their tweets and interactions, we could identify the narratives they promote and their impact on the broader discussion. This information is valuable for understanding the dynamics of opinion formation and dissemination.

\textbf{Engagement Metrics.}
Engagement metrics such as likes, shares, comments, and retweets measured the level of public interest and interaction with tweets related to Generative AI in education. High engagement indicates that a tweet resonates with the audience, sparking discussions and spreading awareness. These metrics helped us identify the most impactful tweets and understand the factors that drive engagement.

\textbf{Temporal Analysis.}
Temporal analysis allowed us to observe how discussions about Generative AI in education evolved over time. By tracking changes in tweet volume, sentiment, and engagement over different periods, we could identify patterns and trends. This analysis helped in predicting future developments and understanding the factors that influence the ebb and flow of public interest and opinion.
Table~\ref{tblInsights} illustrates the insights learnt on the impact of Generative AI on education. The main categories of insights include Personalised Learning, Student Engagement, Career Aspirations, Ethical Concerns, Accessibility, Teacher Support, and Future of Formal Education.

\begin{table}
 \caption{Samples of the insights learnt on the impact of Generative AI on education.}
 \centering
  \begin{tabular}{cc}
   \includegraphics[scale=1.0]{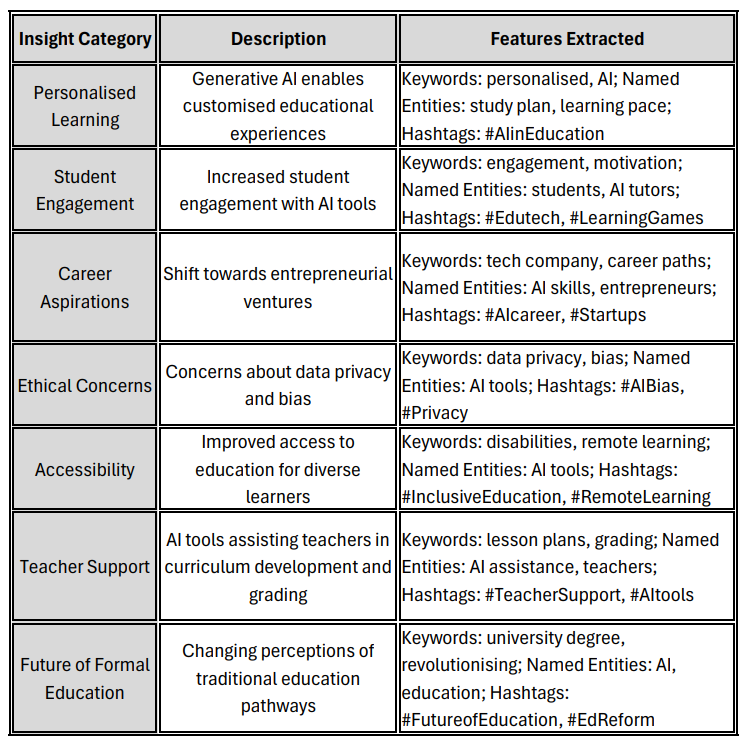}\\
  \end{tabular}
 \label{tblInsights}
\end{table}

\textbf{Personalised Learning.}
Generative AI significantly enhances personalised learning by creating tailored educational experiences that cater to individual student needs. Traditional education systems often struggle to address the diverse learning preferences and paces of students. However, AI systems can adapt learning materials and methods to suit each student. Extracted features such as keywords ("personalised," "AI"), named entities ("study plan," "learning pace"), and hashtags (\#AIinEducation) indicate a high relevance to the domain of personalised learning. These features highlight the customisation capabilities of AI, enabling students to learn at their own pace and focus on areas where they need more practice.

\textbf{Student Engagement.}
Another critical insight is the role of Generative AI in increasing student engagement. AI-driven educational tools, such as interactive learning games and AI tutors, have been found to keep students more engaged and motivated. Extracted features include keywords ("engagement," "motivation"), named entities ("students," "AI tutors"), and hashtags (\#EduTech, \#LearningGames), which emphasise the enhanced interaction and interest levels among students using AI tools. These tools make learning more interactive and enjoyable, helping students stay focused and interested in their studies.

\textbf{Career Aspirations.}
Generative AI is also reshaping students' career aspirations. With access to AI-driven personalised learning, many students are considering entrepreneurial ventures over traditional career paths. The extracted features, such as keywords ("tech company," "career paths"), named entities ("AI skills," "entrepreneurs"), and hashtags (\#AIcareer, \#Startups), reflect this trend. The skills and knowledge gained through AI-enhanced education are seen as valuable assets in the rapidly evolving job market, driving students to pursue careers in AI-related fields.

\textbf{Ethical Concerns.}
Despite the benefits, there are significant ethical concerns regarding the use of Generative AI in education. Issues related to data privacy and algorithmic bias are prominent among students and educators. Extracted features include keywords ("data privacy," "bias"), named entities ("AI tools"), and hashtags (\#AIBias, \#Privacy), highlighting the concerns over how data is used and the potential for biased AI systems. Addressing these ethical concerns is crucial for the responsible integration of AI in education, ensuring that AI systems are fair, transparent, and respect user privacy.

\textbf{Accessibility.}
Generative AI has the potential to improve access to education for diverse learners, including those with disabilities and those in remote areas. AI tools can provide personalised support and resources, making education more inclusive and accessible. Features such as keywords ("disabilities," "remote learning"), named entities ("AI tools"), and hashtags (\#InclusiveEducation, \#RemoteLearning) underscore the importance of AI in bridging educational gaps. These tools help make quality education available to students who might otherwise be underserved.

\textbf{Teacher Support.}
AI tools also assist teachers in various aspects of their work, such as curriculum development and grading. Educators report that AI helps them design better lesson plans and handle grading tasks more efficiently. Extracted features include keywords ("lesson plans," "grading"), named entities ("AI assistance," "teachers"), and hashtags (\#TeacherSupport, \#AItools). This support enhances the overall efficiency and effectiveness of teaching practices, allowing teachers to focus more on personalised student engagement.

\textbf{Future of Formal Education.}
Finally, Generative AI is prompting a reevaluation of traditional education pathways. There is an emerging debate about the necessity of university degrees in an era where AI can provide high-quality, personalised learning experiences. The features extracted, including keywords ("university degree," "revolutionising"), named entities ("AI," "education"), and hashtags (\#FutureofEducation, \#EdReform), highlight the shifting perceptions. This indicates a potential transformation in the structure and delivery of education, with AI playing a central role in shaping the future of formal education.

\subsection{Policy and Regulatory Recommendations}

Based on the findings of this paper, it is evident that Generative Artificial Intelligence (Gen AI) holds transformative potential for the field of education. The integration of Gen AI can lead to significant innovation opportunities, enhancing personalised learning, student engagement, and overall educational outcomes. However, to realise these benefits responsibly, it is crucial to develop comprehensive policies and regulations that support innovation while addressing ethical, societal, and technical challenges. This section outlines key recommendations for policymakers and educational institutions.

\subsubsection{Embracing Innovation Opportunities}

Generative AI offers a wide range of possibilities for innovation in education. By leveraging the capabilities of AI, educators and institutions can enhance teaching and learning experiences, streamline administrative processes, and foster a more inclusive and dynamic educational environment. The following points detail additional innovation opportunities enabled by Generative AI:

Adaptive Learning Systems:
Adaptive learning systems powered by Gen AI can dynamically adjust the difficulty and type of content presented to students based on their individual performance and learning progress. These systems use real-time data to tailor educational experiences to meet the specific needs of each student.
By providing personalised pathways through curriculum material, adaptive learning systems can help students master concepts at their own pace, thereby improving retention rates and academic outcomes.

Intelligent Tutoring Systems:
AI-driven intelligent tutoring systems (ITS) can offer one-on-one tutoring to students, providing personalised feedback, explanations, and hints. These systems simulate the experience of having a personal tutor available around the clock.
ITS can significantly enhance student understanding and performance by offering tailored support and addressing individual learning gaps, which might not be possible in a traditional classroom setting.

Automated Content Creation:
Gen AI can automatically generate educational content such as lesson plans, quizzes, interactive simulations, and multimedia resources. This content creation can be aligned with curriculum standards and customised for different learning styles and levels.
Automated content creation can reduce the workload on educators, allowing them to focus more on direct instruction and student interaction. It also ensures a steady supply of high-quality, diverse educational materials.

Enhanced Collaborative Learning:
AI can facilitate collaborative learning by creating virtual environments where students can work together on projects and assignments. These environments can include AI-mediated discussion forums, group problem-solving activities, and peer review systems.
Collaborative learning supported by AI can foster critical thinking, creativity, and communication skills among students. It also encourages peer-to-peer learning, which can enhance understanding and retention of knowledge.

Real-Time Analytics and Feedback:
AI can provide real-time analytics on student performance, engagement, and progress. Educators can use these insights to make informed decisions about instructional strategies, identify at-risk students, and customise interventions.
Real-time feedback allows for timely adjustments to teaching methods and materials, ensuring that students receive the support they need when they need it most. This proactive approach can lead to better educational outcomes.

Gamification of Learning:
Generative AI can create gamified learning experiences that make education more engaging and enjoyable. This includes developing educational games, interactive challenges, and reward systems that motivate students to achieve learning objectives.
Gamification can increase student motivation and participation, making learning a fun and immersive experience. It can also encourage students to take an active role in their education, fostering a love for learning.

Virtual and Augmented Reality Experiences:
AI can enhance virtual reality (VR) and augmented reality (AR) applications in education, creating immersive learning environments. Students can explore historical events, scientific phenomena, and complex concepts in a virtual space, making abstract ideas tangible and understandable.
VR and AR experiences can deepen understanding and retention by providing hands-on, experiential learning opportunities. They also offer a safe space for experimentation and exploration, which can enhance problem-solving skills and creativity.

Lifelong Learning and Professional Development:
Gen AI can support lifelong learning and continuous professional development by offering personalised learning paths for individuals at different stages of their careers. This includes creating customised courses, recommending relevant resources, and generating new educational content on-demand.
Lifelong learning initiatives supported by AI ensure that individuals can continuously update their skills and knowledge to remain competitive in the job market. This approach promotes a culture of continuous improvement and adaptability.

Language Translation and Multilingual Education:
AI-powered language translation tools can enable multilingual education, allowing students to access learning materials in their native languages. These tools can also assist non-native speakers in understanding and engaging with educational content.
Language translation tools can break down language barriers, making education more inclusive and accessible to students from diverse linguistic backgrounds. This promotes equity and ensures that all students have the opportunity to succeed.

Emotional and Social Learning Support:
AI can monitor and support the emotional and social development of students by analysing behavioral cues and providing appropriate interventions. This includes identifying signs of stress, anxiety, or disengagement and offering resources for emotional well-being.
Emotional and social learning support ensures that students' mental health and well-being are addressed, creating a positive and supportive learning environment. This holistic approach to education fosters overall student success.

\subsubsection{Policy and Regulatory Framework}

To harness the full potential of Generative AI in education while ensuring innovation is pursued with care, the following policy and regulatory recommendations are proposed:

Data Privacy and Security: Policies must prioritise the protection of student data. This includes implementing stringent data privacy regulations, such as data anonymisation, encryption, and access control measures, to safeguard sensitive information. Transparency in data usage and the ability for students and parents to control their data is essential.

Addressing Algorithmic Bias: Policymakers should mandate the development of AI systems that are fair, transparent, and free from biases. Regular audits and evaluations of AI algorithms should be required to detect and mitigate any biases that could perpetuate inequalities in education.

Equitable Access to AI Resources: Ensuring equitable access to AI-driven educational tools is critical. Policies should focus on bridging the digital divide by providing necessary infrastructure, affordable devices, and internet connectivity to all students, regardless of their socioeconomic background.

Ethical AI Use: Establish clear ethical guidelines for the use of AI in education. These guidelines should address issues such as the role of AI in decision-making, the boundaries of AI applications, and the responsibilities of educators and developers in ensuring ethical AI use.

Continuous Professional Development: Invest in continuous professional development for educators to help them integrate AI tools effectively into their teaching practices. Training programs should focus on building AI literacy, understanding AI capabilities and limitations, and fostering a positive attitude towards AI adoption.

Stakeholder Collaboration: Encourage collaboration between policymakers, educational institutions, AI developers, and other stakeholders to create a cohesive strategy for AI integration in education. This collaboration can lead to the development of standards, best practices, and shared resources that benefit all parties.

Monitoring and Evaluation: Implement mechanisms for the ongoing monitoring and evaluation of AI systems in education. This includes assessing the impact of AI tools on learning outcomes, identifying potential risks, and making necessary adjustments to policies and practices.

Support for Innovation: Create a regulatory environment that supports and incentivises innovation in educational technology. This can include funding for research and development, pilot programs to test new AI applications, and public-private partnerships to advance AI-driven educational initiatives.

\subsubsection{Balancing Innovation with Care}

While pursuing the opportunities that Generative AI presents, it is essential to balance innovation with careful consideration of potential risks and ethical implications. Policies should be flexible enough to accommodate rapid technological advancements while ensuring that the primary focus remains on enhancing the educational experience for all students. By fostering a culture of responsible innovation, policymakers and educators can leverage the benefits of Generative AI to create a more personalised, engaging, and equitable educational landscape.


\section{Conclusion and Future Work}

In this paper, we explored the profound impact of Generative Artificial Intelligence (AI) on the future of education, focusing on how it is reshaping educational paradigms and influencing career choices. Our research primarily aimed to understand and predict these impacts by analysing social media discussions using advanced analytical techniques. We developed a comprehensive methodology that leverages ProcessGPT to contextualise social media data and introduced a novel approach for Learning Distributed Representations and Deep Embedded Clustering of Texts. This approach enabled us to extract meaningful insights about the impact of Generative AI on student engagement, personalised learning, career aspirations, and the future of formal education.

The key contributions of this paper include:
(i)~
Development of a Social Media Analysis Framework:
By adapting ProcessGPT technology, we created a robust framework for analysing social media data to understand the impact of Generative AI on education. This framework includes sentiment analysis, topic modeling, user demographic analysis, influencer analysis, engagement metrics, content analysis, hashtag analysis, and temporal analysis;
(ii)~Identification of Emerging Trends:
Our analysis identified and examined emerging trends in using Generative AI in education, such as personalised learning approaches, increased student engagement, and a shift towards entrepreneurial ventures among students;
and (iii)~Policy and Regulatory Recommendations:
Based on our findings, we provided recommendations for policymakers and educational institutions to address challenges related to integrating Generative AI in education, including data privacy, algorithmic bias, and equitable access.
The evaluation of our approach on a Twitter dataset demonstrated the effectiveness of our methodology in accurately contextualising social media data and extracting valuable insights. The high precision, recall, and F1 scores for feature extraction and clustering accuracy validate the robustness of our techniques.

While this paper provides a comprehensive analysis of the impact of Generative AI on education, several areas warrant further research and development. Future work can focus on the following aspects:

\textbf{Personalization Through AI-Enhanced Content Creation:}
As technologies like  Natural Language-Oriented Programming (NLOP)~\cite{beheshti2024natural} emerge, they offer opportunities to integrate natural language-based programming into educational platforms. Future work could investigate how embedding NLOP into AI-driven learning tools might further democratize access to technical education. Research could explore how this paradigm shift would lower the barrier for students with limited programming experience, potentially reducing dropout rates by enhancing engagement and accessibility.
Expanding on current Generative AI capabilities, future research could explore how real-time content adaptation using natural language inputs might cater to diverse learning preferences. Analyzing the interplay between adaptive content delivery and student retention could provide deeper insights into optimizing learning environments for students at risk of churn.

\textbf{Enhanced Data Privacy and Security Measures:}
As data privacy remains a significant concern, future research should explore advanced techniques for ensuring the security and confidentiality of student data used in AI-driven educational applications. This includes developing more robust data anonymisation, encryption, and access control mechanisms.

\textbf{Addressing Algorithmic Bias:} Further work is needed to develop and implement methods for detecting and mitigating algorithmic bias in AI systems. This includes creating transparent and explainable AI models that ensure fairness and equity in educational outcomes~\cite{hanif2023comprehensive,sharma2024explainable}.

\textbf{Expanding Data Sources:} While this research focused on Twitter data, future studies could include data from other social media platforms, educational forums, and online learning platforms. This would provide a more comprehensive understanding of the broader impact of Generative AI on education~\cite{beheshti2021system,luo2024graph}.

\textbf{Longitudinal Studies:} Conducting longitudinal studies to track the long-term effects of Generative AI on education and career choices would provide deeper insights into how these technologies shape learning trajectories and professional pathways over time.

\textbf{Integration with Educational Systems:} Exploring ways to integrate AI-driven tools seamlessly into existing educational systems and curricula is crucial. This includes developing frameworks for teachers and educators to effectively use AI tools in their instructional practices.

\textbf{Evaluation of Learning Outcomes:} Future research should focus on developing new metrics and evaluation frameworks that capture the diverse ways in which students learn and demonstrate knowledge through AI-enhanced education. Traditional metrics may not fully reflect the benefits of personalised and adaptive learning approaches.

\textbf{Ethical and Societal Implications:} Ongoing research should continue to address the ethical and societal implications of using Generative AI in education. This includes examining the broader impacts on educational equity, student well-being, and societal perceptions of education.

\textbf{Developing AI Literacy:} As AI becomes more integrated into education, it is essential to develop AI literacy programs for students, educators, and policymakers. These programs should focus on understanding AI technologies, their potential benefits, and associated risks.


\bibliographystyle{abbrv}
\bibliography{ms}

\end{document}